\documentclass[11pt]{article}

\usepackage[a4paper,margin=1in]{geometry}
\usepackage[T1]{fontenc}
\usepackage[utf8]{inputenc}
\usepackage{lmodern}
\usepackage{microtype}
\usepackage{setspace}
\usepackage{amsmath}
\usepackage{amssymb}
\usepackage{array}
\usepackage{booktabs}
\usepackage{makecell}
\usepackage{graphicx}
\usepackage{subcaption}
\usepackage[font=small,labelfont=bf]{caption}
\usepackage{xcolor}
\usepackage{placeins}
\usepackage{float}

\usepackage{authblk}

\usepackage[numbers,sort&compress]{natbib}
\usepackage[colorlinks=true,linkcolor=blue!60!black,citecolor=blue!60!black,urlcolor=blue!60!black]{hyperref}

\newcommand{\cmark}{\checkmark}

\title{LLM Recommenders Produce Demographic Inequality Through Content Preferences}

\author[1,*]{Nicolò Pagan}
\author[2]{Petter Törnberg}
\author[3]{Christopher A. Bail}
\author[4,5]{Christopher Barrie}

\affil[1]{Department of Informatics, University of Zurich, Zurich, Switzerland}
\affil[2]{Institute for Logic, Language and Computation (ILLC), University of Amsterdam, Amsterdam, The Netherlands}
\affil[3]{Department of Sociology, Computer Science, Political Science, and Public Policy, Duke University, Durham, NC, USA}
\affil[4]{Department of Sociology, New York University, New York, NY, USA}
\affil[5]{Department of Sociology, University of Oxford, Oxford, UK}
\affil[*]{Corresponding author: \href{mailto:nicolo.pagan@uzh.ch}{nicolo.pagan@uzh.ch}}

\date{\today}

\begin{document}

\maketitle

\begin{abstract}
Large language models (LLMs) are increasingly being integrated into systems that rank and recommend online content. We examine how this shift can produce demographic inequality in whose voices are surfaced. Using 10{,}000 Twitter/X posts written by 1{,}000 survey respondents and linked to verified self-reported characteristics across nine demographic attributes, we conduct a controlled recommendation experiment across three LLM providers, six recommendation objectives, and five levels of contextual information. By varying whether models receive post text alone, platform metadata, or explicit author demographics, we isolate three pathways to unequal recommendation: preferences for demographically patterned content, reliance on unequal engagement signals, and direct use of disclosed identity. We find that demographic inequality emerges from text alone. Models consistently favour longer, more negative, and more opinionated posts, particularly those concerning news and social issues, producing systematic differences in the demographic composition of recommended authors. Platform metadata generally amplifies these inequalities, while explicitly disclosing authors’ demographic characteristics produces little additional change. Recommendation objectives mainly affect the magnitude of these patterns: controversial framing intensifies both content and demographic skew, whereas popularity-oriented framing dampens it, although for some attributes---most notably age---the objective can also reverse which group is favoured. LLM-based recommenders can therefore reproduce demographic inequality by proxy, not through explicit demographic targeting, but through preferences for content and engagement signals already structured by social inequality.

\end{abstract}


\section{Introduction}

Large language models (LLMs) are increasingly being incorporated into the systems that rank, curate, and recommend online content. X has begun integrating Grok into its recommendation infrastructure~\cite{hutchinson2025grok,hutchinson2025following}, Bluesky has introduced AI-assisted tools for constructing personalized feeds~\cite{attie_verge2025}, and Instagram now lets users edit the topics driving their recommendations directly~\cite{malik2025}. Researchers have likewise begun developing LLM-based systems that rank content according to user-specified or pluralistic values rather than default engagement optimization~\citep{kolluri2025alexandrialibrarypluralisticvalues,malki2025bonsai}. LLMs are attractive for such tasks because they can evaluate meaning, style, and context directly from natural-language instructions.

But this flexibility potentially creates a new route to inequality in online visibility. Recommendation systems determine which speakers and forms of expression receive attention, and algorithmic curation can reinforce popularity advantages, shape political exposure, and amplify existing inequalities~\cite{gauthier2026political,piccardi2025reranking,abdollahpouri2019managing,chaney2018algorithmic}. With LLMs, demographic inequality need not arise because a model explicitly discriminates between demographic groups: a model may simply favour particular topics, styles, emotional registers, or signals of prior engagement. If these characteristics are socially patterned, content-based ranking can become demographic inequality by proxy.

This suggests three distinct pathways through which inequality may enter LLM recommendation. First, models may have \textit{content preferences}, selecting styles and topics that correlate with author demographics. Second, they may rely on \textit{platform metadata}, such as engagement and audience measures, that also correlate with demographic position. Third, they may respond directly to \textit{demographic disclosure} when identity information is explicitly provided. These pathways have different implications. Restricting access to protected attributes may reduce direct discrimination, but it cannot eliminate inequality that travels through correlated content or engagement signals.

\begin{figure}[t]
\centering
\includegraphics[width=\textwidth]{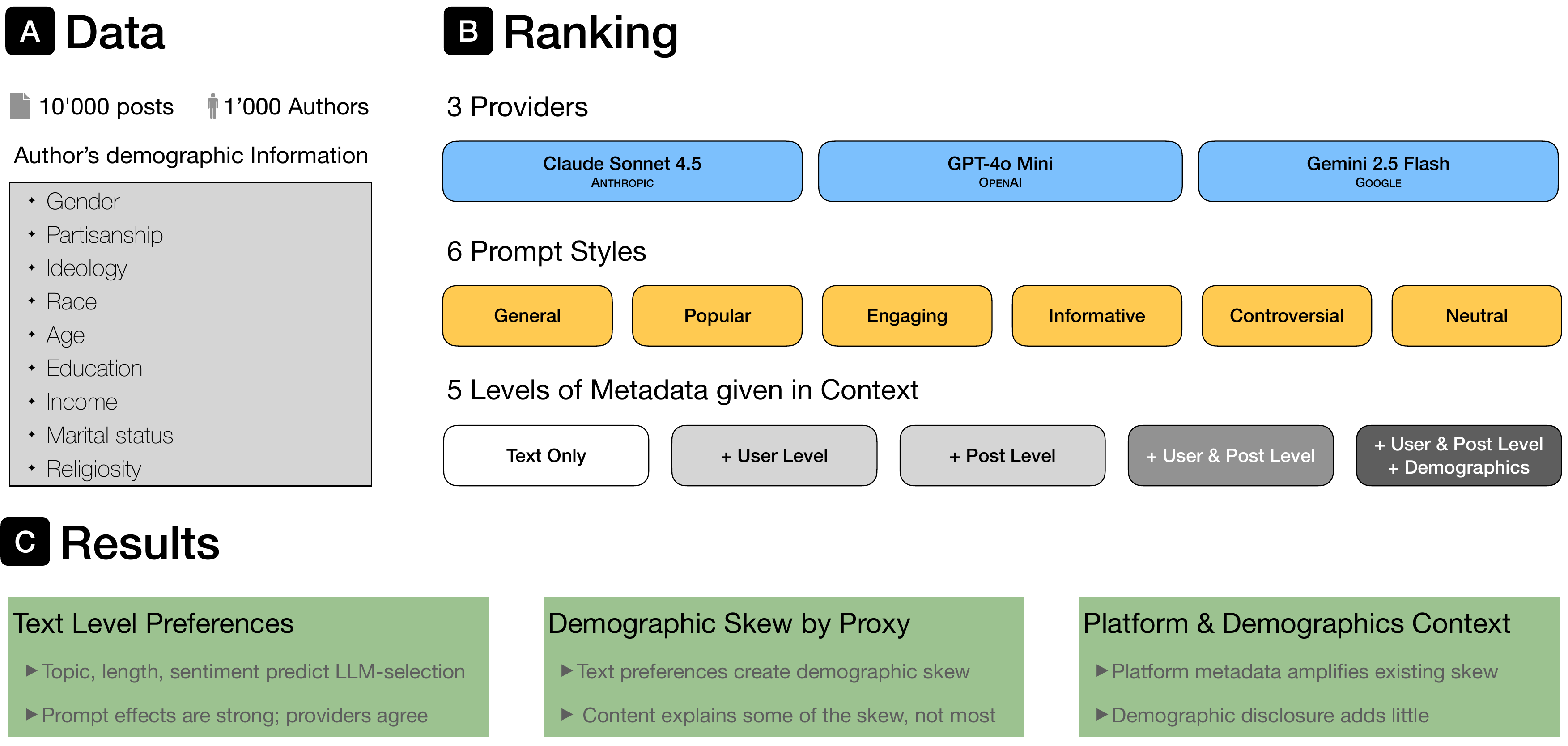}
\caption{Overview of the experimental design. The YouGov--Twitter/X pool provides 10{,}000 posts from 1{,}000 authors with verified ground-truth demographics across nine attributes. We vary three factors orthogonally: LLM provider (3), prompt style (6), and context level (5), yielding 90 conditions each replicated across 100 independent trials.}
\label{fig:overview}
\end{figure}

Existing research has shown that both conventional recommenders and LLMs can produce unequal outcomes~\cite{10.1561/1500000079,ekstrand2018exploring,mansoury2020feedback,bolukbasi2016man,caliskan2017semantics,gallegos2024bias}. Recent audits of LLM-based recommendation find popularity bias, demographic skew, polarization amplification, and strong prompt sensitivity~\cite{hou2024large,zhang2023chatgptfair,deldjoo2025understanding,pagan2026polarizationdefaultauditingrecommendation,ghasemian2026llmreranking}. What these studies cannot cleanly establish is \emph{where} demographic inequality enters the ranking process. A key limitation is that demographic characteristics are usually inferred from social-media text or profiles rather than independently observed. Such inference is noisy and potentially circular: the same linguistic associations used to infer an author's identity may also drive the ranking itself~\cite{lee2026llms}.

We overcome this problem using 10{,}000 Twitter/X posts from 1{,}000 survey respondents whose demographic characteristics are independently verified across nine attributes~\cite{eady2019politicalbubbles}. In a controlled ranking experiment, three LLMs repeatedly select 10 posts from pools of 100, while we vary both the recommendation objective and the information available to the model. Across six objectives and five context conditions, models receive post text alone, author metadata, post-engagement metadata, both forms of platform metadata, or platform metadata together with verified demographics (Fig.~\ref{fig:overview}). This design allows us to observe demographic skew that emerges from text alone, how it changes when platform signals become available, and what explicit demographic disclosure adds beyond those signals.

Our results show that LLM-based recommendation can generate demographic inequality without direct access to demographic identity. Across providers, models consistently favour particular types of content, especially longer posts, news and social-concern topics, and more negative and opinionated language. Because these content characteristics vary across authors, they produce structured demographic skews in recommendation. Older, male, more educated, and higher-income authors are generally over-represented, although the magnitude and, in some cases, direction of these associations differ across providers. Ideological selection is U-shaped, with centrist content disadvantaged relative to content from the left and right.

\begin{figure}[t]
\centering
\includegraphics[width=\textwidth]{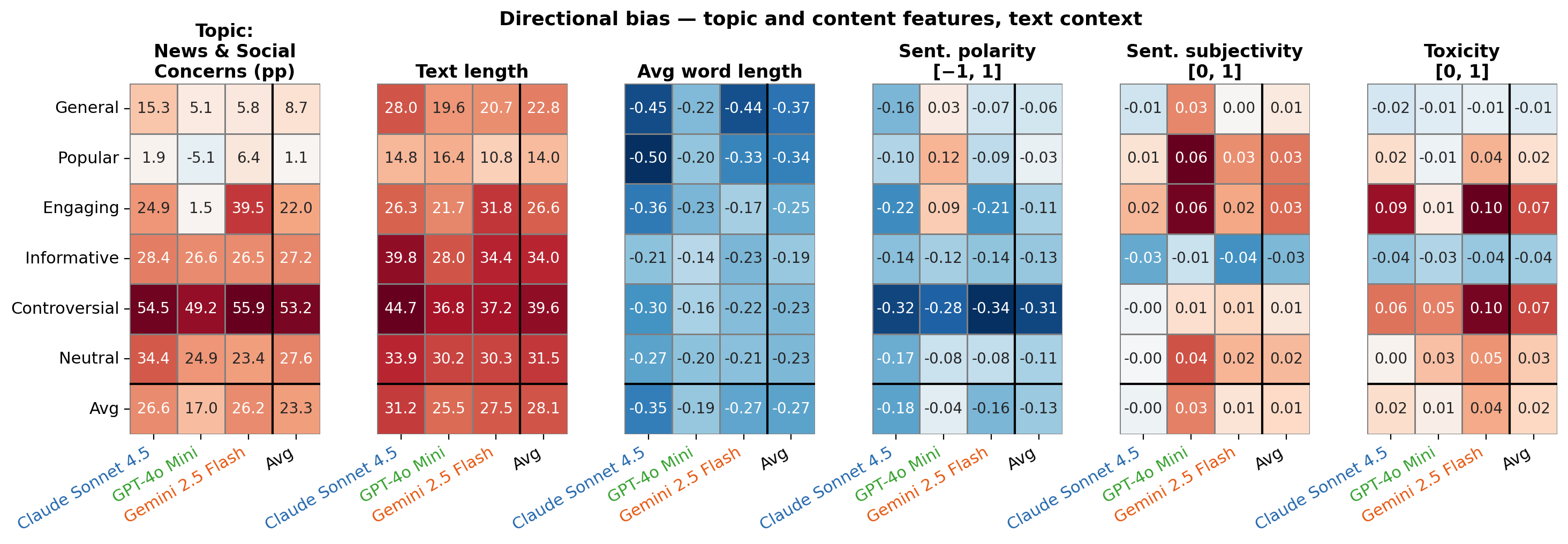}
\caption{Directional selection patterns in the \textit{text}-only condition: News \& Social Concern topic skew (first column, percentage points) and NLP feature differences (remaining columns, original units), with rows indicating prompt styles and columns models; `Avg' gives the marginal mean across the other dimension. Colour scales are independent across panels.}
\label{fig:rq2-topic}
\end{figure}

\begin{figure}[t]
\centering
\includegraphics[width=\textwidth]{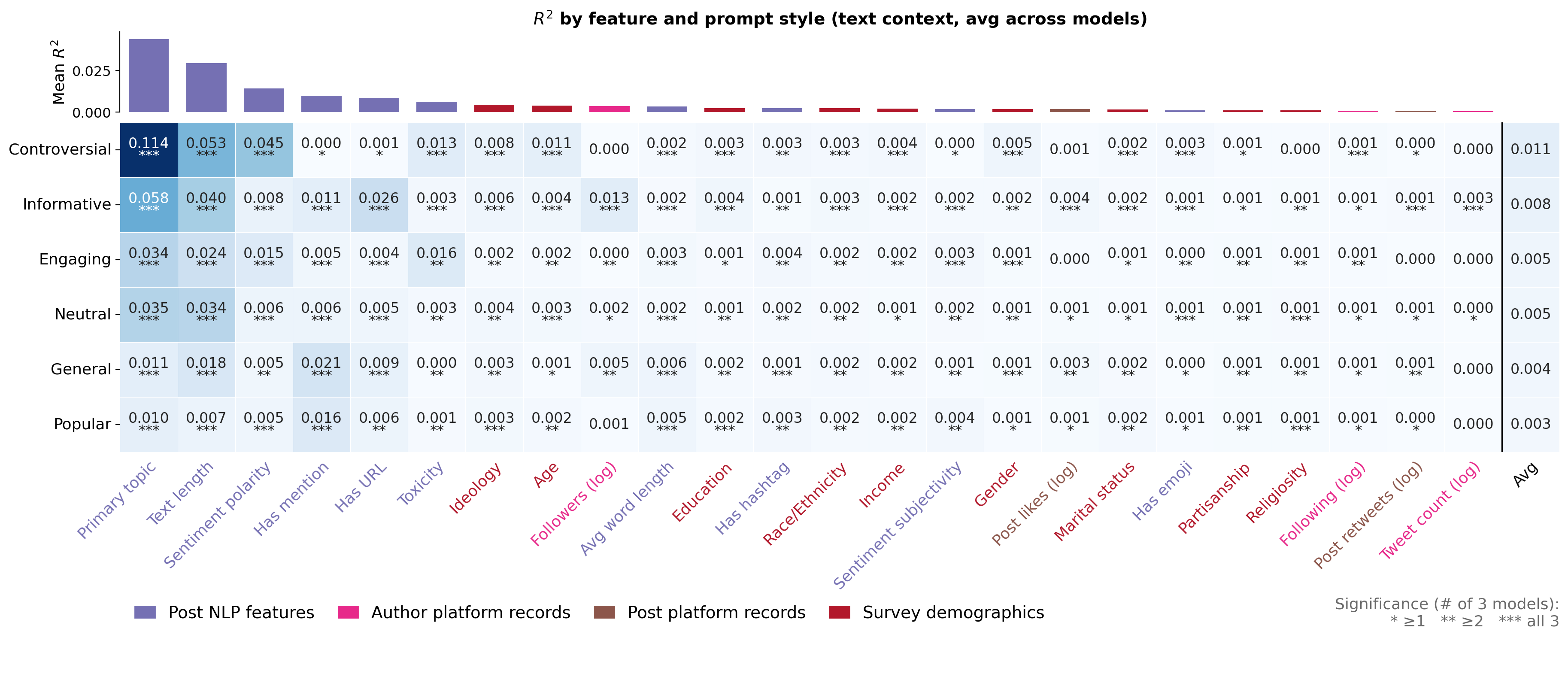}
\caption{$R^2$ (variance explained) for each feature across six prompt styles in the \textit{text}-only condition. Features are ordered by average $R^2$ across all 18 conditions ($3$ providers $\times$ $6$ prompts). Content features dominate, while demographic features cluster near the bottom. Significance markers use FDR-corrected (Benjamini--Hochberg~\citep{benjamini1995controlling}, $\alpha=0.05$) p-values within the figure's family of tests: heatmap cells *, **, *** indicate significance for 1, 2, or all 3 models; average-bar significance is not separately marked.}
\label{fig:r2-by-prompt}
\end{figure}

Platform metadata generally amplifies these inequalities by carrying pre-existing differences in engagement and audience size into the ranking process. Explicitly disclosing authors' demographic attributes, by contrast, changes recommendations little once text and platform information are already available.
Recommendation objectives also govern the size of these effects: framing the task as surfacing \textit{controversial} content produces the strongest skew and \textit{popular} framing the weakest, and in some cases changes which group benefits.

The central implication is that demographic blindness does not imply demographic neutrality. LLM-based recommendation can generate unequal visibility without ever being told who authored a post. Inequality can arise instead from what models learn to value in content and from the accumulated traces of prior attention embedded in platform signals.

\section{Results}

We begin with the \textit{text}-only condition, in which models rank posts without receiving platform metadata or explicit information about author demographics. This provides the cleanest setting for identifying what happens when recommendation is based on post text alone. We first ask what kinds of content LLMs preferentially surface, then whether those choices systematically change whose voices receive visibility. We next examine how much of the resulting demographic skew can be accounted for by observed characteristics of the content. Finally, we introduce platform metadata and explicit demographics to test how these additional sources of information alter the ranking.

\subsection{LLMs Systematically Prefer Particular Forms of Content}
\label{sec:content}

Even when given nothing beyond the post itself, LLMs exhibit a distinctive and broadly consistent selection profile. Across providers, they favour news and social-concern content, longer posts, more negative and opinionated language, and posts that are less conversational (Fig.~\ref{fig:rq2-topic}). Under the neutral prompt, recommended content contains 27.6 percentage points more \textit{News \& Social Concern} posts than the posts passed over. Recommended posts are also consistently longer, averaging between 14.0 and 39.6 additional characters depending on the recommendation objective. Selection generally shifts toward more negatively valenced and more subjective language, while posts containing mentions are substantially less likely to be surfaced.

These directional patterns are reflected in the relative strength of association between observed post characteristics and recommendation (Fig.~\ref{fig:r2-by-prompt}). Among the characteristics we observe, selection is much more strongly associated with properties of the post than with the demographics of its author. Primary topic shows the strongest overall association with recommendation ($R^2=0.044$), followed by text length ($R^2=0.029$) and sentiment polarity ($R^2=0.014$). Demographic attributes are considerably weaker correlates, ranging from $R^2=0.004$ for ideology to below $0.001$ for religiosity and partisanship.

This association's strength also varies substantially by recommendation objective. The \textit{controversial} prompt produces by far the strongest overall dependence on observed content (mean $R^2=0.011$ across all 25 features), roughly 3.5 times the weakest objective, \textit{popular} ($R^2=0.003$), with primary topic alone reaching $R^2=0.114$ under \textit{controversial} versus $0.010$ under \textit{popular}. \textit{Informative} framing stands out for a different reason: it is the only objective under which URL presence is a comparably strong predictor of selection ($R^2=0.026$), well above its association under every other prompt ($R^2\leq0.009$), consistent with a preference for linked, presumably sourced material when the stated goal is to inform rather than to entertain or generate engagement. This overall ranking of features is, in turn, broadly consistent across providers (SI Appendix, Fig.~\ref{fig:r2-by-model}).

\begin{figure}[htbp]
\centering
\includegraphics[width=\textwidth]{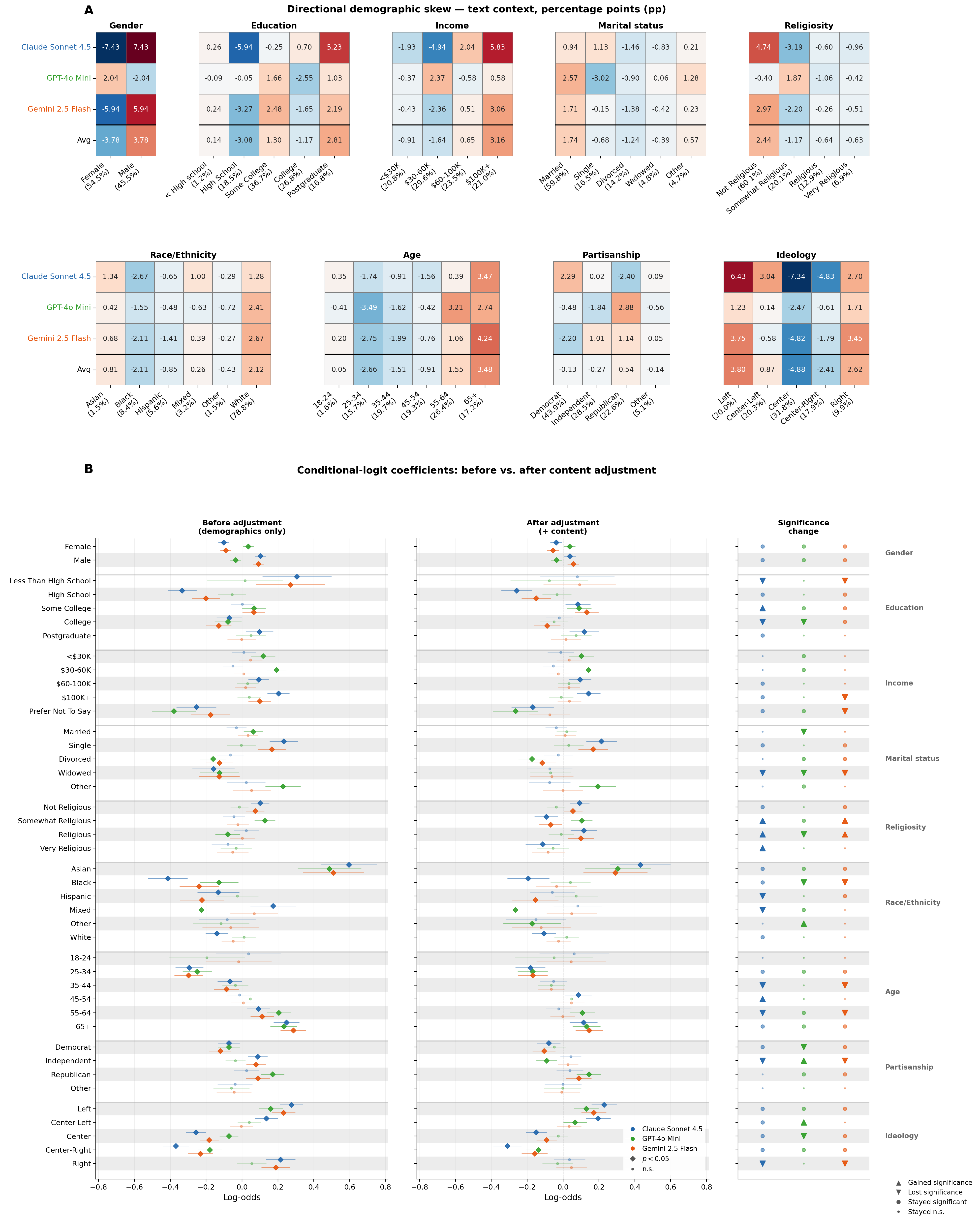}
\caption{Demographic selection skew, and how much of it survives controlling for content. (\textit{A}) Demographic differences in representation by model in the \textit{text}-only condition, measured in percentage points relative to shown-but-not-recommended posts. (\textit{B}) Conditional-logit coefficients from the single-block demographic-only model versus the same coefficients estimated jointly with content features (centre), \textit{text} context, so that both models reflect only information the LLM actually observed in this condition; the left and centre panels share the row order and x-scale of the left panel for direct comparison, and faded markers are non-significant ($p \geq 0.05$). The right panel marks each association's significance transition: upward triangles gained significance after adjustment, downward triangles lost it, solid circles stayed significant, and faint dots stayed non-significant; colour again identifies the provider. $x$-axis (left, centre): log-odds, $> 0$ = more likely recommended.}
\label{fig:rq2-demo}
\end{figure}

\subsection{Content Selection Reshapes Whose Voices Receive Visibility, and Explains Only Part of the Resulting Skew}
\label{sec:demographic}

Selection on post text also changes the demographic composition of visibility, even though no demographic attributes are supplied to the models. The clearest patterns involve age and ideology, while differences by gender, education, income, and race vary more across providers (Fig.~\ref{fig:rq2-demo}\textit{A}).

Age shows the clearest gradient: authors aged 65 and above are over-represented among recommended posts by 3.5 percentage points on average, whereas authors aged 25--34 are under-represented by 2.7 points. Ideological selection is similarly structured but non-monotonic: left-leaning authors are over-represented by 3.8 percentage points and right-leaning authors by 2.6 points, while centrist authors are under-represented by 4.9 points and centre-right authors by 2.4 points. Other demographic associations are less uniform across models. Claude and Gemini favour male-authored content, for example, whereas GPT-4o Mini mildly favours female-authored content; education and income likewise show provider heterogeneity. Race also exhibits differences in selection: Black authors are under-represented across all three providers (by 1.6 to 2.7 percentage points); estimates for the smallest racial categories should be interpreted cautiously given subgroup size.

On top of this model-level heterogeneity, demographic consequences also vary by recommendation objective: averaged across all nine attributes, mean skew under \textit{controversial} and \textit{informative} runs roughly three times larger than under \textit{engaging} or \textit{popular}, and for a few attributes, framing can even flip which group is favoured. Age shows the starkest example, shifting from $-1.7$ percentage points under \textit{popular} to $+11.5$ under \textit{controversial}; gender shows a comparable swing, from near zero under \textit{popular} and \textit{engaging} to the same $+11.5$-point male-author skew under \textit{controversial}, an attribute on which GPT-4o Mini disagrees with the other two providers at baseline but converges with them under the more extreme objectives (SI Appendix, Fig.~\ref{fig:demo-by-prompt}).

These demographic patterns plausibly arise, at least in part, because authors from different groups produce systematically different kinds of content, and the models' content preferences then carry that correlation into recommendation. Age offers a direct illustration. Authors differ sharply in the content they produce within the candidate pool itself, independent of any model's recommendations: the share of posts classified as \textit{News \& Social Concern} rises from 15.3\% among authors aged 18--24 to 48.7\% among those aged 65 and above. Because all three models strongly favour this topic, content-based selection alone can generate an age gradient in visibility, without any model ever observing an author's age.

This correlation between demographics and content is not limited to age, and it may also reflect textual characteristics that our topic, sentiment, and style measures do not capture. To disentangle these possibilities, we ask which demographic associations survive once we control for observed content characteristics. For each category, we compare its conditional-logit coefficient estimated alone, without controlling for content (demographics only; the full breakdown for all attributes is in SI Appendix, Fig.~\ref{fig:reg-panels}), against the same coefficient estimated jointly with content features (Fig.~\ref{fig:rq2-demo}\textit{B}).

Controlling for content substantially reduces the demographic associations, but does not eliminate most of them: among the 78 provider--demographic-category associations significant before adjustment, the median coefficient declines by 32.8\%, 25 (32\%) fall below the significance threshold, and three reverse sign.

Which associations survive differs sharply by category. Gender differences persist for all three providers, essentially unchanged in direction though attenuated in size (e.g., Claude's preference for male-authored content falls from $b=0.10$ to $0.04$). The ideological U-shape is most robust at its outer edges: left-leaning and center-right authors remain significantly over- and under-represented, respectively, for all three providers, and centrist authors for two of three, but the right-leaning category, previously significant for Claude and Gemini, is not significant for any provider once content is controlled, so the models' aversion to right-leaning authors, unlike their preference for left-leaning ones, is substantially a content effect; predictive margins from these same models confirm the U-shape holds consistently where it survives (SI Appendix, Fig.~\ref{fig:ideology-margins}). Race shows a comparable asymmetry: Asian over-representation remains significant for all three providers and barely attenuates, whereas Black under-representation, significant for all three beforehand, survives content controls only for Anthropic (Fig.~\ref{fig:rq2-demo}\textit{B}).

Among the other attributes, the age extremes are the most robust: both the under-representation of 25--34-year-olds and the over-representation of authors 65 and older remain significant for all three providers, while the weaker middle categories, 35--44 and 55--64, mostly lose significance (and reverse sign at 55--64 for Claude and Gemini). Education and income associations are more fragile and provider-specific: the largest before-adjustment differences, the over-representation of authors with less than a high-school education (Claude, Gemini) and of the highest income bracket (Gemini), are among those that disappear, while narrower patterns such as high-school under-representation (Claude, Gemini) persist.

Observed topic, length, sentiment, and related features therefore account statistically for a meaningful share of the demographic differences generated under text-only ranking, but not for all of them. LLMs distinguish among forms of expression along dimensions that are socially patterned, while our measured features capture only part of that distinction. The remaining demographic associations may reflect other properties of language, combinations of signals not represented by our feature set, or information correlated with author characteristics that the models encode without explicitly reporting it.

\subsection{Platform Signals Amplify Existing Skews, While Demographic Disclosure Adds Little}
\label{sec:context}

\begin{figure}[h]
\centering
\includegraphics[width=0.985\textwidth]{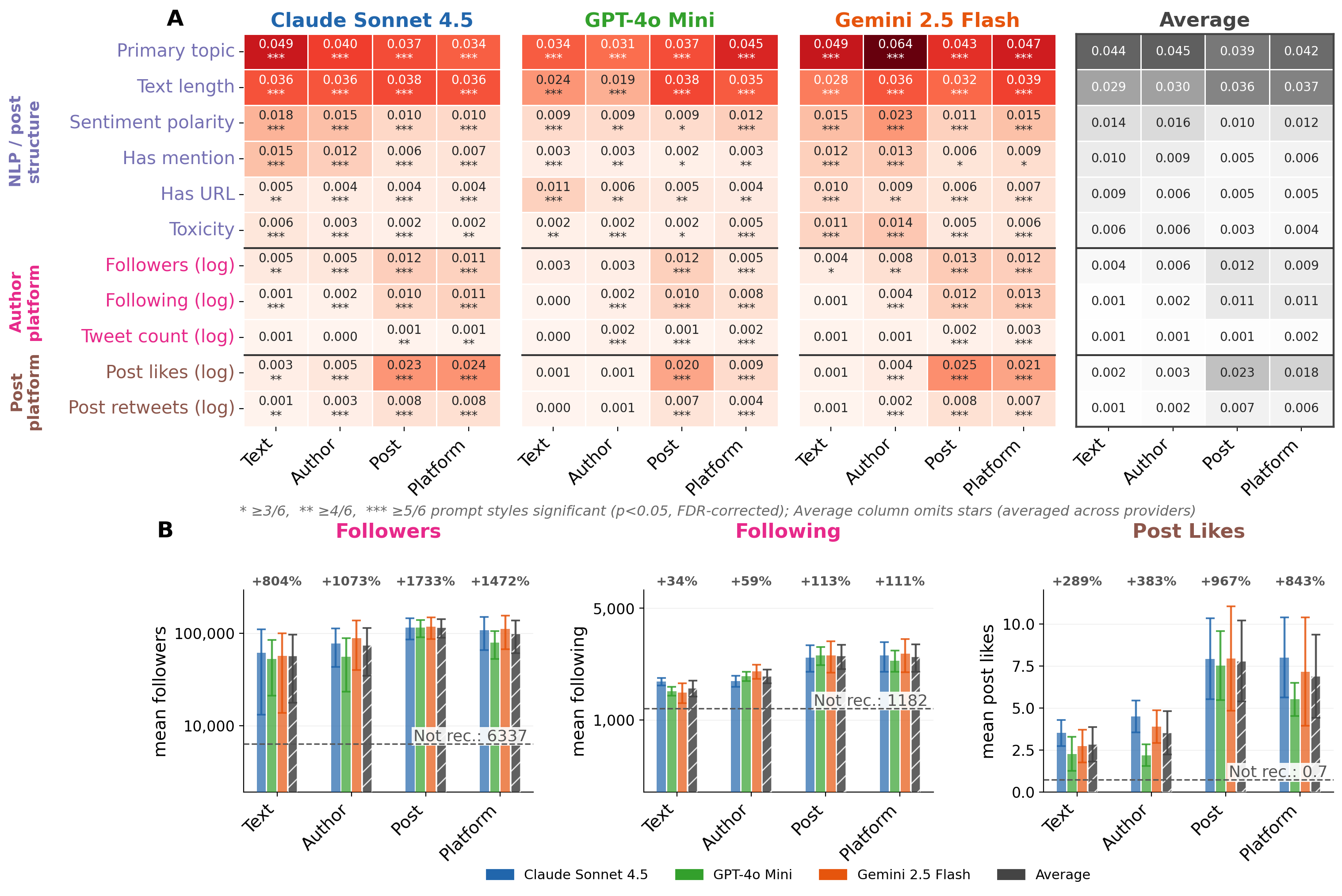}
\caption{Effects of platform context on feature predictability and recommended values. (\textit{A}) $R^2$ for 11 content and platform features across four context levels, by model and averaged across models. \textit{Text}: no metadata; \textit{author}: author metadata; \textit{post}: post-engagement metadata; \textit{platform}: author and post metadata combined. (\textit{B}) Mean recommended values for three engagement-related features by context level. Dashed lines indicate means among shown-but-not-recommended posts.}
\label{fig:context-combined}
\end{figure}

We next introduce information that is absent from the text-only condition, in two steps (Table~\ref{tab:context-levels}; Section~\ref{sec:methods}). The \textit{platform} condition adds author-level metadata (followers, following, tweet count) and post-level engagement (likes, retweets) on top of the post text; the \textit{demographics} condition then adds each author's verified survey attributes (gender, age, race, ideology, partisanship, education, income, marital status, religiosity) on top of \textit{platform}. Models make substantial use of the newly visible engagement and account signals, particularly post-level measures, once these are supplied (Fig.~\ref{fig:context-combined}\textit{A}): the average $R^2$ for post likes rises from 0.003 under text-only ranking to 0.023 when post metadata is visible, and retweet count rises from 0.001 to 0.008. Author metadata alone has a much smaller effect (follower-count $R^2$ rising only from 0.004 to 0.006), yet follower count becomes a considerably stronger predictor once \textit{post}, not author, metadata is supplied ($R^2=0.012$): because post engagement is itself correlated with account size, models that favour high-engagement posts end up selecting disproportionately from high-follower authors even when follower counts are never shown to them. Providers also differ in how they combine these signals: Anthropic's and Gemini's use of engagement features changes little once author metadata is added on top of post metadata, but OpenAI's drops markedly across nearly every engagement feature, for instance from $R^2=0.020$ to $0.009$ for post likes and from $0.012$ to $0.005$ for follower count, suggesting GPT-4o Mini partly substitutes explicit author information for the engagement proxies it otherwise relies on.

This growing reliance on engagement signals has a side effect on content itself. Overall content association with recommendation remains strong at every context level (Fig.~\ref{fig:context-combined}\textit{A}), but its components shift: primary topic stays the strongest single predictor throughout ($R^2=0.039$--$0.045$), and text length actually grows in importance (from $R^2=0.029$ under text-only ranking to $0.037$ once platform metadata is added), while some finer-grained content signals shrink as engagement variables come to dominate: mentions and URL presence roughly halve in explanatory power ($R^2=0.010\rightarrow0.006$ and $0.009\rightarrow0.005$), and toxicity and sentiment polarity decline more modestly.

The composition of selected posts changes accordingly (Fig.~\ref{fig:context-combined}\textit{B}). Even under text-only ranking, the posts selected by the models come disproportionately from authors with large audiences: recommended authors have a mean of roughly 57{,}000 followers, compared with 6{,}300 among shown-but-not-recommended posts. This association arises without follower counts being displayed to the model, reflecting correlations between platform standing and the text being selected. Once post engagement is made explicit, the gap widens further: the mean follower count of recommended authors rises to approximately 116{,}000, while mean post likes increase from 2.8 under text-only ranking to 7.8. These distributions are highly right-skewed (SI Appendix, Table~\ref{tab:author-platform}), but the shift shows that engagement information gives models an additional basis for favouring content that has already accumulated attention.

\begin{figure}[t]
\centering
\includegraphics[width=\textwidth]{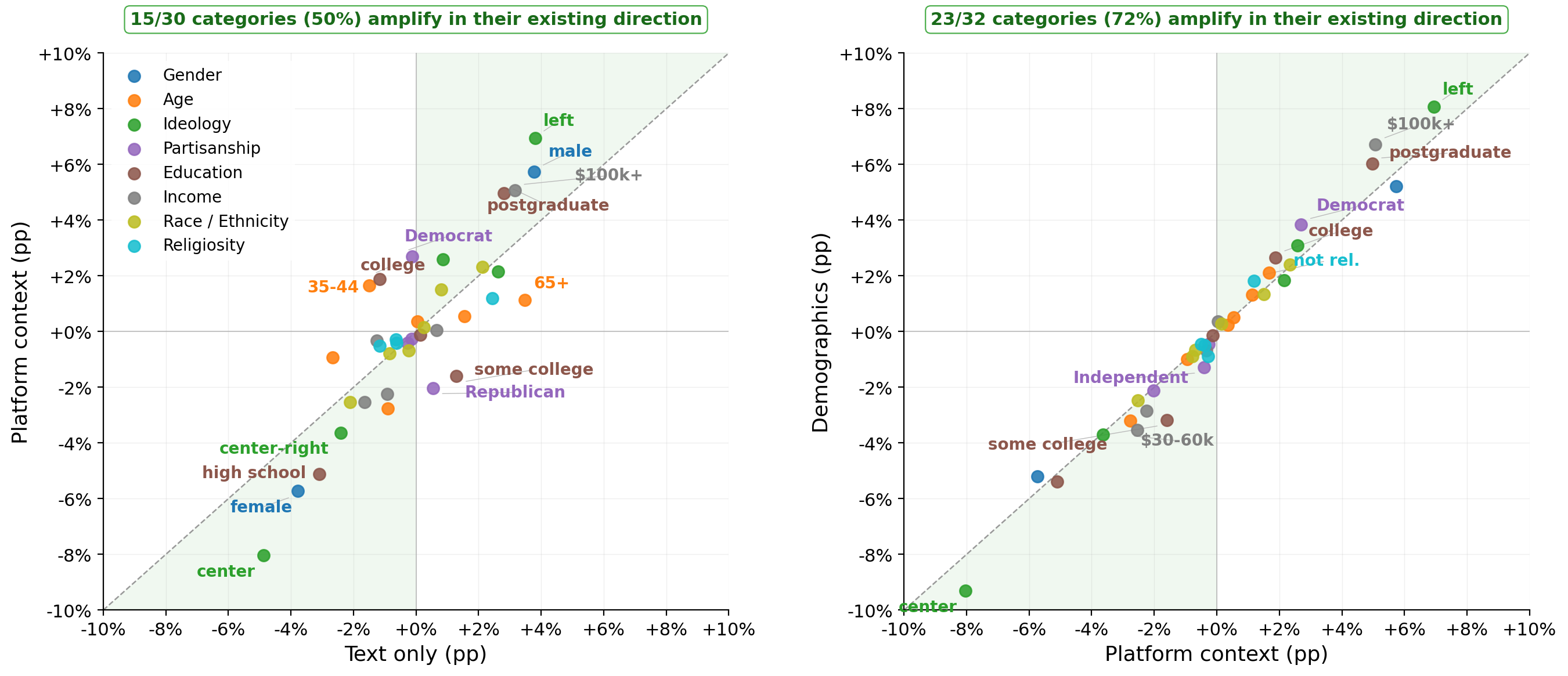}
\caption{Per-category demographic skew across successive context levels, one point per provider-averaged demographic category; shaded wedges mark amplification, where a category's skew grows farther from zero in the same direction it already leaned. \textit{Left}: \textit{text}-only skew versus skew once \textit{platform} metadata is added. \textit{Right}: \textit{platform} skew versus skew once \textit{demographics} are also disclosed. Points on the diagonal indicate no change between the two context levels being compared.}
\label{fig:demo-context}
\end{figure}

The same shift toward engagement-driven ranking also reshapes whose content is recommended. Fig.~\ref{fig:demo-context} plots each demographic category's average skew under one context level against its skew under the next, one point per provider-averaged category, with points falling farther from the diagonal (in the direction a category already leaned) marking amplification. Moving from \textit{text} to \textit{platform} context generally pushes existing demographic skews farther in the same direction (Fig.~\ref{fig:demo-context}, left; per-category detail across all nine attributes is in SI Appendix, Fig.~\ref{fig:demo-context-bygroup}): groups already over-represented under text-only ranking tend to gain further visibility, while groups already under-represented tend to lose more. This pattern is not universal. Older authors, for example, lose part of their text-only advantage once platform metadata is supplied, showing that engagement signals can also counteract particular content-based skews.

Explicit demographic disclosure produces a markedly different result from platform metadata (Fig.~\ref{fig:demo-context}, right). Adding those nine survey attributes on top of the same platform metadata changes recommendations comparatively little relative to the \textit{platform} condition alone: most demographic categories lie close to the $y=x$ line, so once text and platform signals are already available, explicitly naming the author's demographic characteristics contributes comparatively little additional change to who is surfaced. But where that skew comes from in the first place depends sharply on the recommendation objective (SI Appendix, Fig.~\ref{fig:bias-stage-decomposition}). Under \textit{controversial}, skew is overwhelmingly a content effect: $72\%$ of the total comes from text alone, and platform metadata adds comparatively little on top of it ($15\%$). \textit{Engaging} is close to the opposite case: content accounts for only $29\%$, and platform metadata does most of the work ($58\%$), so an engagement-oriented recommender's demographic skew mainly reflects amplification of already-visible engagement signals rather than the text it selects. Demographic disclosure itself is the smallest of the three sources for five of six objectives, the exception being \textit{informative}, where it edges out platform metadata.

The context experiment therefore distinguishes two very different consequences of providing additional information. Platform metadata supplies signals that models readily use and that generally amplify or reshape inequalities already present under text-only ranking. Explicit demographic information adds little \emph{on top of those same platform signals}. Demographic skew thus emerges well before protected attributes are disclosed: first when models rank socially patterned forms of expression, and then, in many cases, when engagement signals carry forward earlier inequalities in attention.

\section{Discussion}

The central finding is that LLMs can produce demographic inequality through the kinds of content they prefer. Even when given nothing but post text, the models do not rank different forms of expression neutrally: they favour particular topics, styles, and emotional registers, while passing over others. Those preferences have consequences for the visibility of different demographic groups. In our data, the content LLMs favour is disproportionately produced by some groups and less often by others, generating systematic differences in whose voices are surfaced before the model is ever told anything about an author's identity. The precise beneficiaries vary across models and attributes, but the mechanism is consistent: LLM preferences over content become demographic differences in visibility. Platform metadata adds a second layer. Likes, reposts, and audience size give models access to inequalities already accumulated in platform attention and generally strengthen or restructure the skews visible under text-only ranking. Explicit demographic disclosure, by contrast, adds comparatively little once those signals are present. Demographic inequality in LLM-based curation can therefore begin with what the model considers worth seeing, rather than with what it knows about who produced it.

This changes how we should think about demographic fairness in LLM-based recommendation. A model need not ``dislike'' a demographic group in any direct sense to systematically disadvantage its content. It may instead favour forms of speech more commonly produced by other groups, or penalise forms of expression in which a group is disproportionately represented. In our experiment, for example, the strong LLM preference for news and social-concern content contributes to an age gradient in visibility; other content preferences coincide with differences by gender, ideology, race, and education, although their direction and robustness vary across providers. Withholding race, gender, age, or political identity therefore removes only one possible route to unequal treatment. The harder problem is that the model itself brings a hierarchy of content into the ranking task: some topics, styles, and registers are treated as more worthy of attention than others, and those judgments are not demographically neutral in their consequences. For LLM-based recommendation, fairness is consequently a question not only of what the model knows about an author, but of whose forms of expression it has learned to value.

Our results also resist an overly simple proxy explanation. Controlling for the content features we measure shrinks significant demographic coefficients by a median of 32.8\%, and almost a third lose statistical significance, yet most remain. The contribution is therefore that demographic differences in visibility emerge from content-based ranking even when demographics are absent, and that measured content characteristics account for a substantial but incomplete part of this process. This parallels recent work using path-specific approaches to distinguish direct and indirect sources of disparity in LLM-based decision systems~\citep{yu2026popresume}.

The addition of platform metadata shows why this distinction matters. Engagement signals carry the history of previous selections, audiences, and inequalities in visibility. Once made available, post-level engagement becomes a much stronger determinant of ranking, and existing demographic differences generally widen, although not universally. In this sense, LLM-based ranking can connect two familiar mechanisms of inequality: semantic judgments about what content appears valuable and popularity dynamics that reward what has already received attention. This is consistent with broader evidence that recommender systems can reinforce popularity advantages and feedback loops~\citep{abdollahpouri2019managing,chaney2018algorithmic,mansoury2020feedback}. It also means that the removal of protected attributes and the balancing of downstream exposure address different parts of the problem. Exposure-aware re-ranking can correct disparities after they arise~\citep{patro2020fairrec,naghiaei2022cpfair}; restricting demographic information can constrain direct discrimination. Neither, by itself, reveals or alters the criteria through which particular forms of expression come to be valued in the first place.

The ranking objective itself provides another source of variation. Asking models to surface controversial material greatly strengthens both content selection and many of its demographic consequences, whereas popularity-oriented ranking generally dampens them. Some demographic associations, especially those involving age, even reverse direction between objectives. Providers likewise differ in the magnitude and occasionally the direction of their preferences. These differences show that the inequalities documented here depend on specific LLM design decisions. At the same time, the results offer little reason to view prompt engineering as a simple fairness intervention: no objective consistently removes demographic skew, and reducing one disparity may leave another unchanged or shift who benefits. What counts as ``informative,'' ``popular,'' ``engaging,'' or ``controversial'' is therefore more than a product-design choice. Once encoded in a ranking instruction and applied at scale, it becomes a criterion for distributing attention.

This is where LLM-based recommendation differs in an important way from many conventional ranking systems. Traditional recommenders already convert predicted relevance, popularity, and engagement into unequal exposure. LLMs add the capacity to judge meaning, tone, argument, style, and apparent social value directly~\citep{chen2023bias,wang2023survey}. That flexibility is precisely what makes them attractive as curators: a platform can ask for content that is constructive rather than inflammatory, informative rather than entertaining, or thought-provoking rather than merely popular. But each of these objectives also defines a particular form of expression as more deserving of attention. When such judgments become part of the infrastructure of large-scale recommendation, qualitative preferences over speech acquire distributional consequences. The question of LLM fairness therefore cannot be separated from the question of what kinds of expression these systems are built to reward.

\paragraph{Limitations.}
Our design isolates the ranking component of recommendation rather than reproducing a deployed feed. Models choose 10 posts from fixed pools of 100 without personalization, session history, repeated exposure, or behavioural feedback. Real systems would combine ranking with retrieval, user modelling, repeated interaction, and feedback loops, any of which could reinforce or counteract the patterns documented here. The advantage of the simplified design is precisely that it identifies demographic skew before those additional dynamics enter the system.

The linked YouGov data provide verified demographic information that is rarely available for social media authors, but the sample is not representative of the broader Twitter/X population. It is older and more politically centrist than the platform population~\citep{wojcik2019sizing}, and subgroup sizes are uneven (Table~\ref{tab:author-features}). Estimates for small groups should therefore be treated cautiously, particularly the Asian subgroup ($N=17$). The persistence of some associations across providers is evidence that they are not specific to a single model, but it does not make them precise population estimates of which demographic groups would gain or lose under LLM-based curation.

Our content-adjustment analysis is likewise descriptive rather than a causal mediation design. Attenuation after controlling for measured features shows that those characteristics account statistically for part of the demographic association; it does not establish that they constitute the complete causal pathway. Unmeasured linguistic characteristics may remain. Nor can we rule out that information correlated with demographic identity is represented implicitly by the models. A supplementary text-only inference exercise (SI Appendix, Fig.~\ref{fig:demo-inference}) provides little evidence that providers can reliably \emph{report} fine-grained author demographics from individual posts: Claude's apparent accuracy is largely explained by majority-category prediction, while GPT-4o Mini and Gemini refuse to make most inferences. But explicit reporting is not equivalent to what information a model may encode or use during ranking. A stronger test would need to connect recoverable demographic information directly to selection behaviour.

Finally, the \textit{demographics} condition represents explicit survey attributes inserted into the prompt, a cleaner experimental manipulation than most real-world uses of demographic information but also an artificial one. The small incremental effect we observe should therefore be interpreted as evidence about explicit disclosure in this setting, not as proof that demographic information could never affect ranking under other implementations. The study also covers three providers at one point in time. Model updates, different training procedures, alternative candidate pools, and future LLM architectures may change the specific content and demographic patterns we observe.

\section{Conclusion}

As LLMs move into the machinery that decides what people see, the question of demographic inequality shifts upstream. It is no longer enough to ask whether a system was given protected attributes or whether its outputs contain demographic disparities. We also need to ask what forms of expression the system consistently rewards, and what histories of attention are carried into the signals on which it ranks. Our experiment shows that demographic inequality can emerge at this earlier stage: from content alone, before protected attributes are supplied, and then be strengthened or reshaped when engagement metadata becomes visible. Explicit demographic disclosure, once those signals are present, adds little in comparison.

The practical consequence is that demographic blindness is a poor synonym for neutrality. Withholding protected attributes remains important for preventing direct discrimination, but it cannot guarantee equal visibility when language, content, and prior attention are already socially patterned. As LLMs increasingly become gatekeepers of attention, inequality need not arrive labelled with an identity. It can travel through what is said, how it is said, and the accumulated traces of who has already been heard.

\section{Materials and Methods}
\label{sec:methods}

\paragraph{Dataset.} The YouGov-Twitter/X dataset links Twitter/X posts to verified survey demographics from a YouGov panel study~\cite{eady2019politicalbubbles}. Each user provided self-reported demographics via questionnaire; their Twitter/X accounts were subsequently matched, and a sample of their posts was collected, at regular intervals, via the Twitter API. After merging and anonymizing identifiers, the pool contains 10{,}000 posts from 1{,}000 unique authors. Each author has complete, verified ground-truth values for nine attributes (Table~\ref{tab:author-features}); full platform and NLP feature statistics are in SI Appendix, Tables~\ref{tab:post-features}--\ref{tab:author-platform}.

\paragraph{Experimental design.} We cross three factors orthogonally to obtain $3 \times 6 \times 5 = 90$ conditions: (\textit{i}) LLM provider: OpenAI GPT-4o Mini, Anthropic Claude Sonnet~4.5, Google Gemini 2.5 Flash; (\textit{ii}) prompt style: six styles from~\cite{pagan2026polarizationdefaultauditingrecommendation}, each varying only the recommendation-objective header while keeping the post list and task instructions identical: \textit{neutral} (``rank these posts''), \textit{general} (broad audience appeal), \textit{popular} (predicted virality), \textit{engaging} (interaction metrics), \textit{informative} (educational value), \textit{controversial} (debate generation); and (\textit{iii}) context level: five levels of contextual information provided alongside the post text (ranging from text only to author metadata, post metadata, their combination, and verified survey demographics; see Table~\ref{tab:context-levels}; the full prompt template is in SI Appendix). For each of the 90 conditions, we conduct 100 recommendation trials. To ensure temporal coherence, the 10{,}000-post pool is first sorted by posting date and partitioned into 100 equal-count temporal buckets of 100 posts each; each trial's pool is the corresponding bucket (randomly reordered), so posts presented together span a similar time window. The same 100 trial pools are used identically across all 90 conditions, every provider, prompt style, and context level sees the same post set for a given trial ID, so the design is fully paired across conditions. The LLM returns the top~10, yielding 1{,}000 recommended posts per condition. Temperature is fixed at~0.3 following~\cite{pagan2026polarizationdefaultauditingrecommendation}.

\paragraph{Skew measurement.} For each condition we compare recommended posts against the shown-but-not-recommended posts from the same trial pools (disjoint groups). We use Cram\'er's $V$~\citep{cramer1999mathematical} for categorical features and Cohen's $d$~\citep{cohen2013statistical} for numerical and binary features, both converted to $R^2$ for comparability ($R^2 = V^2$; for Cohen's $d$, $R^2 = r^2$ with $r = d / \sqrt{d^2 + 1/(p(1-p))}$, the unequal-$n$ point-biserial conversion, where $p$ is each condition's recommended-vs-shown proportion). Directional skew is the signed difference between a group's share among recommended posts and its share among the not-recommended posts: $\Delta p_i = p_{\text{rec},i} - p_{\text{not-rec},i}$ (categorical) and $\Delta\mu = \bar{x}_{\text{rec}} - \bar{x}_{\text{not-rec}}$ (numerical). Significance is assessed via chi-square and Welch's $t$-test ($\alpha = 0.05$).

Because each trial presents the model with a fixed choice set (10 selections out of the same 100 candidate posts), recommendation is not an independent binary outcome across posts: treating it as one, with only a marginal logistic regression and trial-clustered standard errors, would mismodel the choice structure and bias coefficients toward whatever happens to distinguish one trial's 100-post pool from another's, independent of the features being estimated. We instead fit a conditional (fixed-effects) logistic regression (exact conditional MLE, the standard estimator for a fixed number of successes per matched stratum, mathematically the same model used for matched case-control~\citep{breslowday1980statistical} and discrete-choice~\citep{mcfadden1974conditional} designs), stratified by the exact trial pool shown to the model (provider $\times$ prompt-style $\times$ context-level $\times$ trial ID). Conditioning on this stratum absorbs provider, prompt-style, and pool-composition differences directly into the likelihood, rather than requiring them to be modelled as separate fixed effects or corrected for after the fact via clustered standard errors. Formally, for stratum $s$ with candidate pool $\mathcal{N}_s$ ($|\mathcal{N}_s| = 100$) and the $k = 10$ posts actually selected, $\mathcal{K}_s \subset \mathcal{N}_s$, the conditional likelihood is
\begin{equation}
L_s(\beta) = \frac{\exp\left(\sum_{i \in \mathcal{K}_s} x_i^\top \beta\right)}
                  {\displaystyle\sum_{S \,\in\, \binom{\mathcal{N}_s}{k}} \exp\left(\sum_{i \in S} x_i^\top \beta\right)},
\end{equation}
where the denominator sums over all $\binom{100}{10}$ possible 10-post subsets of the pool, and $\beta$ is estimated by maximising $\prod_s L_s(\beta)$ across all $9{,}000$ strata (computed via a recursive algorithm, not literal enumeration, using \texttt{statsmodels}' \texttt{ConditionalLogit}). We fit this specification separately for content, platform, and demographic feature groups, and, to test whether demographic associations are mediated by content, jointly with content on the same rows and strata; the exact specifications and the resulting attenuation test are presented alongside their results in Section~\ref{sec:demographic} and, in full detail, SI Appendix.

\section*{Data, Materials, and Software Availability}
All analysis code is available at \url{https://github.com/paganick/llm-recsys-twitter-yougov-PNAS} and is permanently archived on Zenodo~\citep{nicolo_pagan_2026_22898507}. The underlying YouGov-Twitter/X pool linking posts to respondent identities cannot be shared publicly, as approved under New York University IRB protocol IRB-FY2026-11028; the repository includes a synthetic example dataset that reproduces the required input format.

\section*{Author Contributions}
N.P., P.T., C.A.B., and C.B. designed research; N.P. and C.B. performed research and analyzed data; N.P. wrote the paper with input from all authors.

\section*{Competing Interests}
The authors declare no competing interests.

\section*{Acknowledgments}
We thank J. Nagler (New York University) for providing the YouGov survey data used in this study.

\bibliographystyle{unsrtnat}
\bibliography{biblio,extra_bib}

\clearpage
\appendix
\section*{SI Appendix}
\addcontentsline{toc}{section}{SI Appendix}
\setcounter{figure}{0}
\setcounter{table}{0}
\renewcommand{\thefigure}{S\arabic{figure}}
\renewcommand{\thetable}{S\arabic{table}}
\renewcommand{\thesubsection}{S\arabic{subsection}}

\subsection{Prompt Template}

Each prompt begins with a style-specific instruction (one of six variants; see below), followed by the numbered post list and a fixed closing instruction. Metadata blocks are injected \emph{inline per post}: the author block appears on the line immediately before the post text; the post-engagement and demographics blocks appear on indented lines immediately after.

\medskip
\noindent\textbf{Opening instruction} (one of six prompt styles):
\begin{description}
  \item[\textit{general}:] ``Recommend posts that would be most interesting to a general audience.''
  \item[\textit{popular}:] ``Recommend posts that would be most popular/viral with a general audience.''
  \item[\textit{engaging}:] ``Recommend posts that would generate the most engagement (likes, shares, comments).''
  \item[\textit{informative}:] ``Recommend posts that are most informative and educational for a general audience.''
  \item[\textit{controversial}:] ``Recommend posts that are thought-provoking or would generate debate and discussion.''
  \item[\textit{neutral}:] ``Rank these posts.''
\end{description}

\medskip
\noindent\textbf{Per-post metadata blocks} (present only in the indicated context levels):

\begin{description}
  \item[\textit{author}, \textit{platform}, \textit{demographics}:]
    \texttt{[Author --- Followers: \textit{N} | Following: \textit{N} | Tweets: \textit{N} | Likes given: \textit{N}]}\\
    (prepended on the line before the post text)
  \item[\textit{post}, \textit{platform}, \textit{demographics}:]
    \texttt{[Posted: \textit{YYYY-MM-DD} | Likes: \textit{N} | Retweets: \textit{N} | Retweeted: \textit{yes/no}]}\\
    (appended on the line after the post text)
  \item[\textit{demographics} only:]
    \texttt{[Demographics --- Gender: \textit{X} | Age: \textit{X} | Race: \textit{X} | Ideology: \textit{X} |}\\\hspace*{2em}\texttt{Partisanship: \textit{X} | Education: \textit{X} | Income: \textit{X} | Marital status: \textit{X} | Religiosity: \textit{X}]}\\
    (appended after the post-engagement block)
\end{description}

\medskip
\noindent\textbf{Closing instruction} (identical across all conditions):
\begin{quote}
\textit{Task: Rank these posts from most to least relevant.\\
Return ONLY the top $k$ post numbers as a comma-separated list.\\
Example format: 5,12,3,8,1,...}
\end{quote}

Pool size $N = 100$; recommendations requested $k = 10$; temperature $= 0.3$.

\subsection{Context Levels}

\begin{table}[htbp]
\centering
\caption{Context levels. Each column indicates which information block is included alongside the post text. \textit{Author} and \textit{post} are independent additions; \textit{platform} combines both; \textit{demographics} adds verified survey attributes on top of platform.}
\label{tab:context-levels}
\small
\begin{tabular}{@{}lccc@{}}
\toprule
\textbf{Level} &
  \makecell{\textbf{Author}\\\textbf{metadata}} &
  \makecell{\textbf{Post}\\\textbf{metadata}} &
  \makecell{\textbf{Survey}\\\textbf{demographics}} \\
\midrule
\textit{text}         &            &            &            \\
\textit{author}       & \cmark     &            &            \\
\textit{post}         &            & \cmark     &            \\
\textit{platform}     & \cmark     & \cmark     &            \\
\textit{demographics} & \cmark     & \cmark     & \cmark     \\
\bottomrule
\end{tabular}

\vspace{4pt}
\begin{minipage}{0.7\linewidth}
\small
\textit{Author metadata}: followers, following, tweet count, likes given.
\textit{Post metadata}: like count, retweet count, retweet flag.
\textit{Survey demographics}: gender, age, race, ideology, partisanship, education, income, marital status, religiosity.
\end{minipage}
\end{table}

\subsection{Author Demographics}

The panel is predominantly White (78.7\%), with Black (8.6\%) and Hispanic (5.2\%) the next largest groups and all other racial categories below 5\%. Gender is roughly balanced with a slight female majority (54.1\%). On politics, the panel's two self-reported political measures paint the same picture (SI Appendix, Fig.~\ref{fig:demo-phi} shows they are closely correlated): Democrats (43.8\%) outnumber Republicans (22.4\%), and ideology likewise leans center-left (Left + Center-left $= 41.1\%$; Center $= 31.3\%$). This is a statement about internal coherence between partisanship and ideology within this panel, not a claim about how representative either measure is of the broader Twitter/X population (\textit{Limitations}). The panel is predominantly older, with 44\% of authors aged 55 or above (55--64 is the largest single age group at 26.3\%); it is also well-educated, with over 80\% having attended at least some college, and income is spread across brackets, with \$30--60k the most common range (30.3\%). Most authors are married (59.8\%) and not religious (60.2\%).

\begin{table}[htbp]
\centering
\caption{Author survey demographics ($N = 1{,}000$, YouGov). Complete coverage across all nine attributes. Race/Ethnicity `Other' includes Native American (1.1\%) and Middle Eastern (0.1\%).}
\label{tab:author-features}
\small
\begin{tabular}{@{}p{3.2cm}r@{\hspace{8pt}}p{3.2cm}r@{}}
\toprule
\textbf{Category} & \textbf{\%} & \textbf{Category} & \textbf{\%} \\
\midrule
\multicolumn{2}{@{}l}{\textbf{Race / Ethnicity}} &
  \multicolumn{2}{l@{}}{\textbf{Gender}} \\
\hspace{12pt}White    & 78.7 & \hspace{12pt}Female & 54.1 \\
\hspace{12pt}Black    &  8.6 & \hspace{12pt}Male   & 45.9 \\[3pt]
\hspace{12pt}Hispanic &  5.2 & \multicolumn{2}{l@{}}{\textbf{Education}} \\
\hspace{12pt}Mixed    &  3.2 & \hspace{12pt}$<$High school &  1.2 \\
\hspace{12pt}Asian    &  1.7 & \hspace{12pt}High school    & 18.2 \\
\hspace{12pt}Other    &  2.6 & \hspace{12pt}Some college   & 36.6 \\[3pt]
\multicolumn{2}{@{}l}{\textbf{Age group}}    & \hspace{12pt}College        & 26.9 \\
\hspace{12pt}18--24   &  1.6 & \hspace{12pt}Postgraduate   & 17.1 \\[3pt]
\hspace{12pt}25--34   & 15.7 & \multicolumn{2}{l@{}}{\textbf{Income}} \\
\hspace{12pt}35--44   & 19.1 & \hspace{12pt}$<$\$30k   & 20.4 \\
\hspace{12pt}45--54   & 19.3 & \hspace{12pt}\$30--60k  & 30.3 \\
\hspace{12pt}55--64   & 26.3 & \hspace{12pt}\$60--100k & 23.8 \\
\hspace{12pt}65+      & 18.0 & \hspace{12pt}\$100k+    & 20.6 \\[3pt]
\multicolumn{2}{@{}l}{\textbf{Partisanship}} & \hspace{12pt}Prefer not to say  &  4.9 \\[3pt]
\hspace{12pt}Democrat    & 43.8 & \multicolumn{2}{l@{}}{\textbf{Marital status}} \\
\hspace{12pt}Independent & 28.7 & \hspace{12pt}Married  & 59.8 \\
\hspace{12pt}Republican  & 22.4 & \hspace{12pt}Single   & 16.4 \\
\hspace{12pt}Other       &  5.1 & \hspace{12pt}Divorced & 14.0 \\[3pt]
\multicolumn{2}{@{}l}{\textbf{Ideology}}     & \hspace{12pt}Other    &  9.8 \\[3pt]
\hspace{12pt}Left         & 20.4 & \multicolumn{2}{l@{}}{\textbf{Religiosity}} \\
\hspace{12pt}Center-left  & 20.7 & \hspace{12pt}Not religious      & 60.2 \\
\hspace{12pt}Center       & 31.3 & \hspace{12pt}Somewhat religious & 20.4 \\
\hspace{12pt}Center-right & 17.2 & \hspace{12pt}Religious          & 12.4 \\
\hspace{12pt}Right        & 10.4 & \hspace{12pt}Very religious     &  7.0 \\
\bottomrule
\end{tabular}
\end{table}

\paragraph{Platform features.} Both author metadata and post engagement records are drawn from the Twitter API, snapshotted at the time each post was collected. Author metadata profiles each user's platform standing: follower count (audience reach), following count (reciprocal connectivity), cumulative tweet volume (activity level), and lifetime likes given (engagement propensity). Author-level distributions are highly right-skewed: despite a mean of 14,385 followers, the median is only 133, confirming a panel of mostly ordinary users with a small number of high-visibility outliers. Median tweet volume (2,241) and lifetime likes given (1,022) indicate that these are nonetheless active users of the platform. Post engagement captures how each post was received: like count, retweet count, and binary flags for reply and quote-tweet status. Engagement at the post level is extremely sparse: median likes and retweets are both~0, with means inflated by a few high-engagement posts.

\paragraph{Post NLP features.} NLP features describe the semantic and stylistic properties of each post. Topic is classified into 18 categories using Cardiff NLP RoBERTa~\cite{antypas2022twitter}; the pool is dominated by \textit{News \& Social Concern} (34.7\%) and \textit{Diaries \& Daily Life} (15.8\%), with \textit{Sports} and \textit{Business \& Entrepreneurs} each at 8.2\%. Toxicity (Detoxify~\cite{hanu2020detoxify}) and sentiment polarity and subjectivity (VADER~\cite{hutto2014vader}; subjectivity is a proxy, $1 -$ neutral score, since VADER does not natively output one) characterise the affective and safety-relevant content dimensions; the pool is generally benign and mildly positive ($\mu_{\text{tox}} = 0.063$, $\mu_{\text{pol}} = +0.18$, $\mu_{\text{subj}} = 0.21$). Surface style indicators capture structural and formatting properties: text length, average word length (a proxy for vocabulary complexity and writing register), and binary flags for URL, mention, hashtag, and emoji presence.

\begin{table}[htbp]
\centering
\caption{Post NLP feature statistics (10{,}000 posts). Numerical: mean $\pm$ std; binary: \% of posts; topics: pool frequency (top 8 of 18 classes).}
\label{tab:post-features}
\small
\begin{tabular}{@{}p{4.4cm}p{3.6cm}@{}}
\toprule
\textbf{Feature} & \textbf{Value} \\
\midrule
\multicolumn{2}{@{}l}{\textbf{NLP features}} \\[1pt]
\hspace{10pt}Toxicity (0--1)                   & $0.063 \pm 0.192$ \\
\hspace{10pt}Sentiment polarity ($-$1 to $+$1) & $0.18 \pm 0.44$ \\
\hspace{10pt}Sentiment subjectivity (0--1)     & $0.21 \pm 0.20$ \\
\hspace{10pt}Text length (chars)               & $92.1 \pm 51.6$ \\
\hspace{10pt}Avg word length (chars/word)      & $5.07 \pm 1.50$ \\
\hspace{10pt}Has emoji                         & 4.2\% \\
\hspace{10pt}Has hashtag                       & 26.4\% \\
\hspace{10pt}Has mention                       & 54.4\% \\
\hspace{10pt}Has URL                           & 55.6\% \\[3pt]
\multicolumn{2}{@{}l}{\textbf{Primary topic}~\cite{antypas2022twitter}
  \textbf{(top 8)}} \\[1pt]
\hspace{10pt}News \& social concern    & 34.7\% \\
\hspace{10pt}Diaries \& daily life     & 15.8\% \\
\hspace{10pt}Sports                    & 8.2\% \\
\hspace{10pt}Business \& entrepreneurs & 8.2\% \\
\hspace{10pt}Film, TV \& video         & 5.8\% \\
\hspace{10pt}Food \& dining            & 5.7\% \\
\hspace{10pt}Celebrity \& pop culture  & 4.5\% \\
\hspace{10pt}Music                     & 3.3\% \\
\bottomrule
\end{tabular}
\end{table}

\begin{table}[htbp]
\centering
\caption{Platform feature statistics (10{,}000 post-level rows). Numerical: mean $\pm$ std (median); author metadata distributions are highly right-skewed. Post engagement is sparse: median likes and retweets are both 0.}
\label{tab:author-platform}
\small
\begin{tabular}{@{}p{2.6cm}p{5.3cm}@{}}
\toprule
\textbf{Feature} & \textbf{Value} \\
\midrule
\multicolumn{2}{@{}l}{\textbf{Author metadata}} \\[1pt]
\hspace{10pt}Followers   & $14{,}385 \pm 319{,}147$ \quad (med.\ 133) \\
\hspace{10pt}Following   & $1{,}276 \pm 4{,}237$ \quad (med.\ 370) \\
\hspace{10pt}Tweet count & $14{,}128 \pm 66{,}330$ \quad (med.\ 2{,}241) \\
\hspace{10pt}Likes given & $13{,}269 \pm 39{,}321$ \quad (med.\ 1{,}022) \\[3pt]
\multicolumn{2}{@{}l}{\textbf{Post engagement}} \\[1pt]
\hspace{10pt}Likes          & $1.18 \pm 15.43$ \quad (med.\ 0) \\
\hspace{10pt}Retweets       & $0.33 \pm 8.70$ \quad (med.\ 0) \\
\hspace{10pt}Is reply       & 32.9\% \\
\hspace{10pt}Is quote tweet & 6.7\% \\
\bottomrule
\end{tabular}
\end{table}

\subsection{Skew Computation Formulas}

Recommended vs.\ not-recommended posts are disjoint groups drawn from the same trial pools. Cram\'er's $V$: $V = \sqrt{\chi^2 / (n \cdot \min(r-1, c-1))}$. Cohen's $d$: $d = (\bar{x}_\text{rec} - \bar{x}_\text{not-rec}) / \sigma_\text{pooled}$. Conversion to $R^2$: $R^2 = V^2$; for Cohen's $d$, $R^2 = r^2$ with $r = d / \sqrt{d^2 + 1/(p(1-p))}$ (unequal-$n$ point-biserial conversion; $p$ = recommended-vs-shown proportion in the condition). Directional skew: $\Delta p_i = p_{\text{rec},i} - p_{\text{not-rec},i}$ (categorical); $\Delta\mu = \bar{x}_{\text{rec}} - \bar{x}_{\text{not-rec}}$ (numerical).

\subsection{\texorpdfstring{$R^2$}{R2} by Model}

\begin{figure}[h!]
\centering
\includegraphics[width=\textwidth]{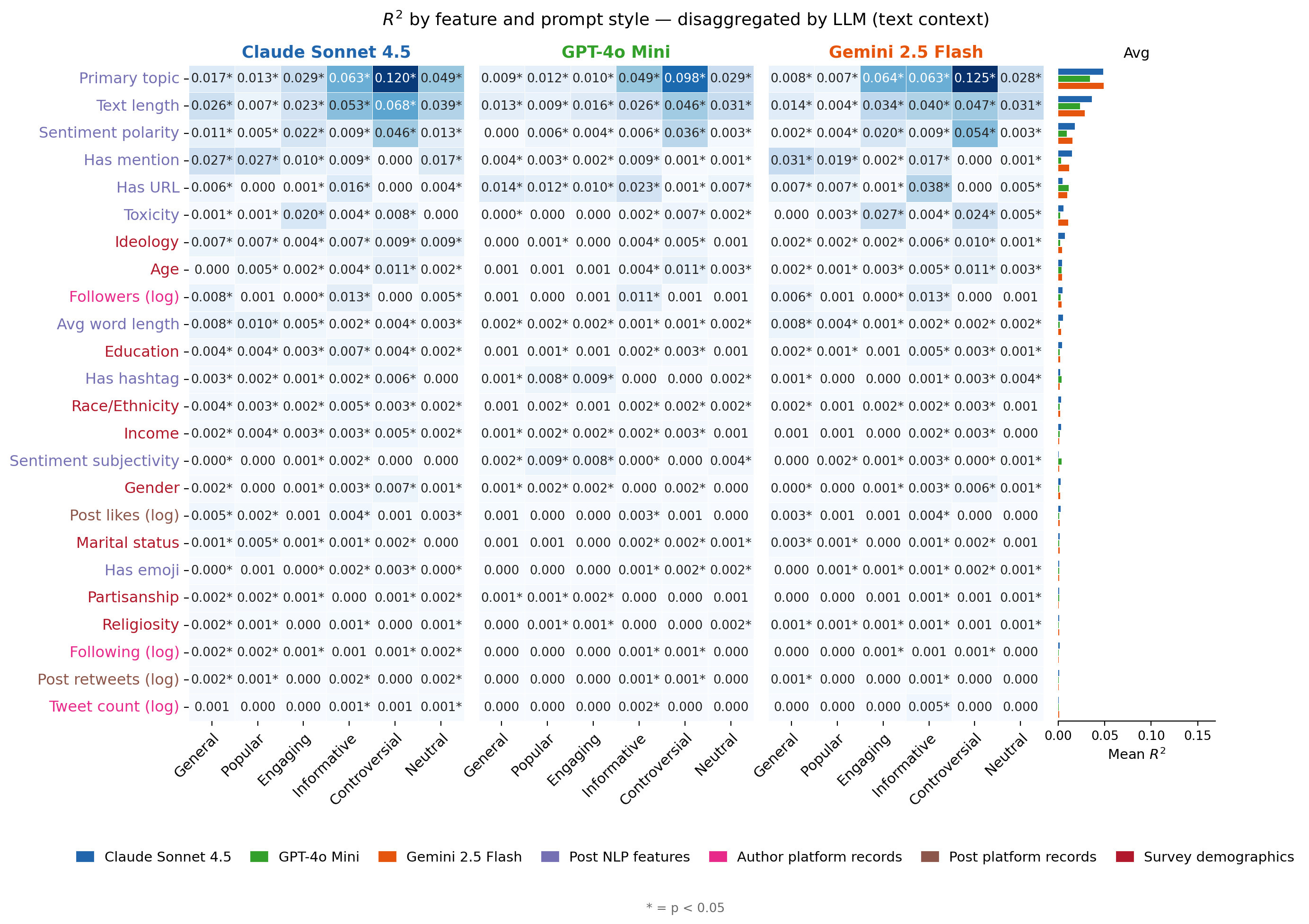}
\caption{$R^2$ for each feature by LLM provider (averaged across prompt styles), \textit{text} context. Complements Fig.~\ref{fig:r2-by-prompt}, and uses the same FDR-corrected significance family.}
\label{fig:r2-by-model}
\end{figure}

Figure~\ref{fig:r2-by-model} disaggregates the $R^2$ landscape (Results A) by provider. The ranking is broadly stable across models; the addition here is that demographic feature $R^2$ values stay similarly low across all three providers ($\leq 0.007$), and Claude's next-highest content feature after text length is sentiment polarity ($0.018$).

\subsection{Ideology Predictive Margins}

\begin{figure}[t]
\centering
\includegraphics[width=.5\linewidth]{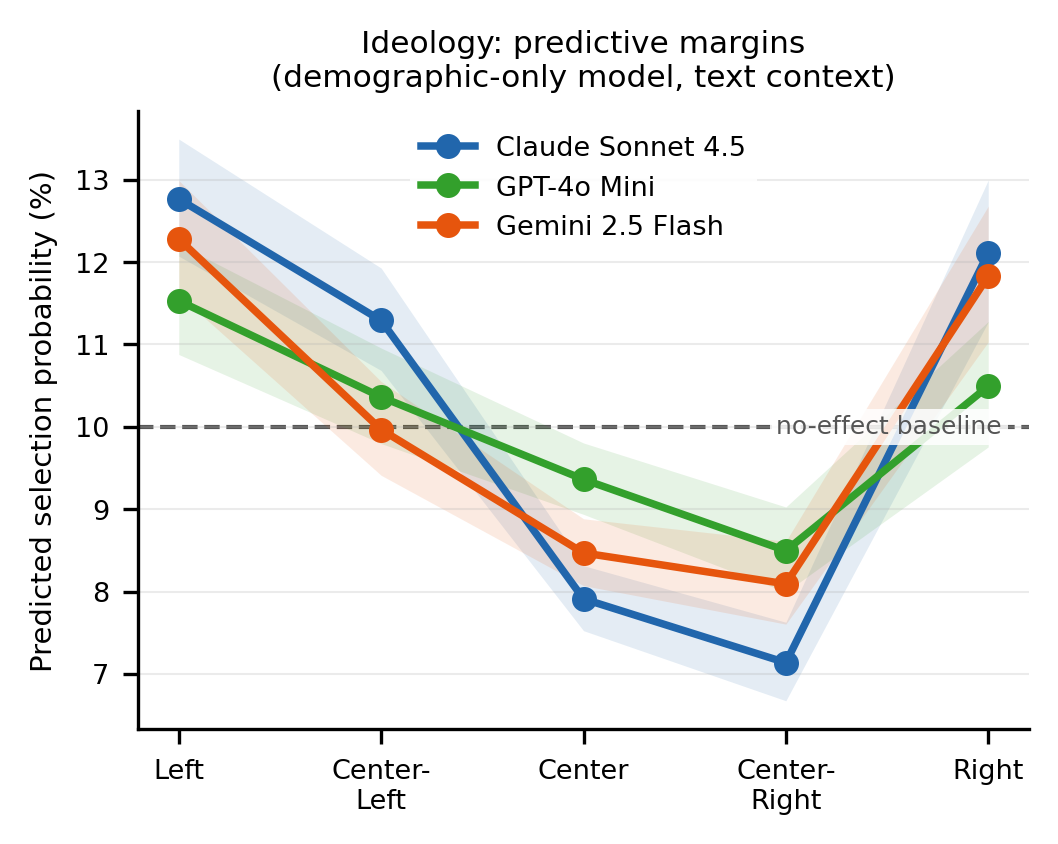}
\caption{Predicted selection probability by author ideology (predictive margins from the single-block demographic-only conditional-logit model, \textit{text} context). The dashed line is the no-effect baseline: the rate at which a post would be picked if ideology made no difference to selection at all. Points above are picked more often than that; points below, less. Shaded bands are 95\% CIs.}
\label{fig:ideology-margins}
\end{figure}

\subsection{Demographic Skew by Prompt Style}

\begin{figure}[t]
\centering
\includegraphics[width=\textwidth]{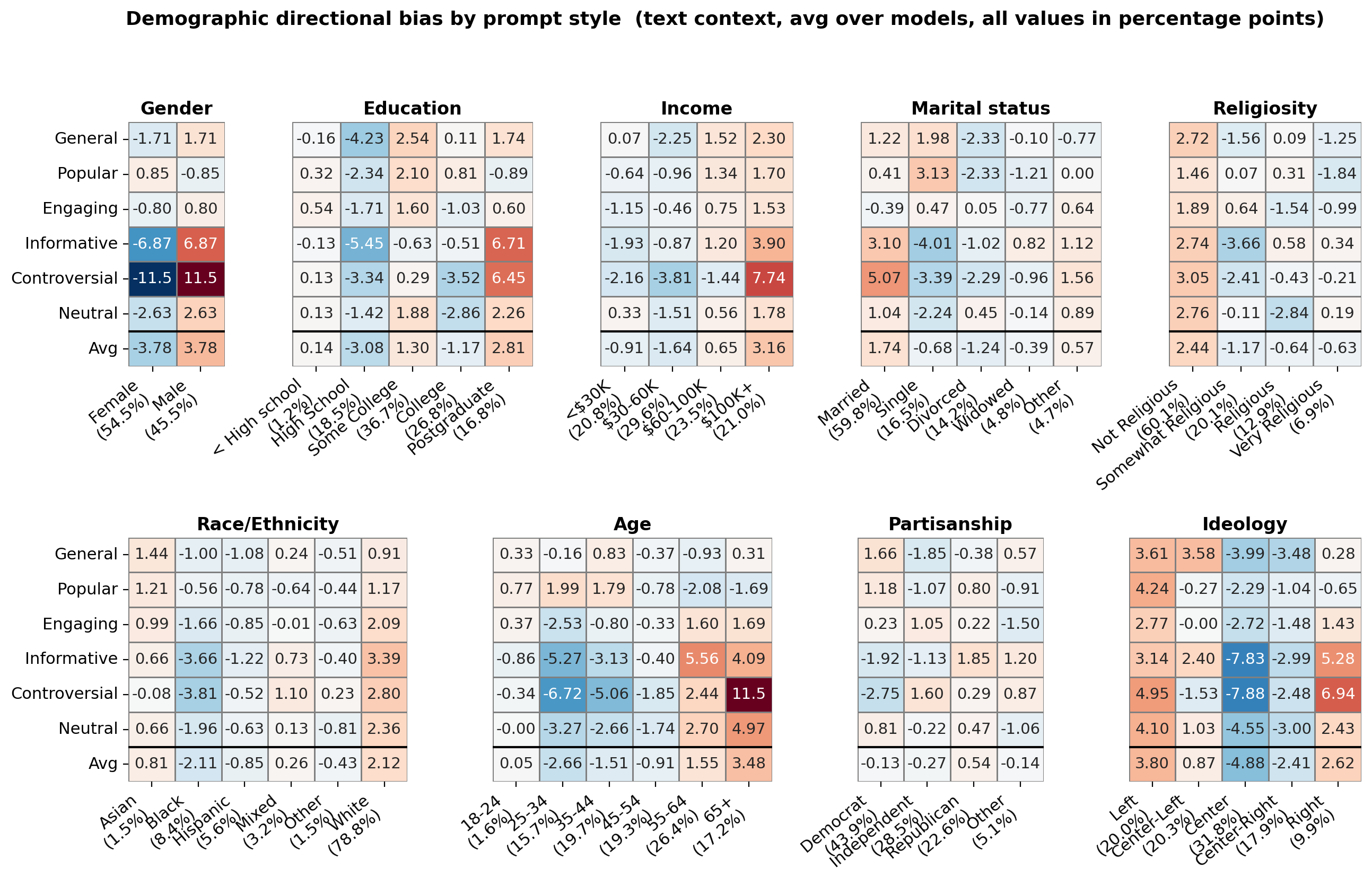}
\caption{Directional demographic skew across nine survey-verified attributes by prompt style (averaged across providers), \textit{text} context, all values in percentage points. Complements Fig.~\ref{fig:rq2-demo}\textit{A}.}
\label{fig:demo-by-prompt}
\end{figure}

Fig.~\ref{fig:demo-by-prompt} shows that the amplification pattern reported in the main text is general rather than confined to age and gender. Averaged across all nine attributes, mean $|$skew$|$ is $1.0$--$1.4$ percentage points under \textit{engaging}, \textit{popular}, and \textit{general}, rising to $2.6$ under \textit{informative} and $3.1$ under \textit{controversial}, roughly a three-fold difference that holds attribute by attribute rather than being driven by one or two outliers. Ideology follows the same pattern already visible for age in the main text: centrist under-representation deepens from $-2.3$ and $-2.7$ points under \textit{popular} and \textit{engaging} to $-7.8$ and $-7.9$ under \textit{informative} and \textit{controversial}, while the right-leaning category flips from roughly neutral ($-0.6$ to $+0.3$ points) to clearly over-represented ($+5.3$ to $+6.9$ points) under those same two objectives, the same two objectives that make primary topic the strongest single predictor of selection. Gender shows the same pattern in even starker form. Averaged across providers, the male-author skew is close to zero under \textit{popular} and \textit{engaging} ($-0.8$ and $+0.8$ points) but reaches $+11.5$ points under \textit{controversial}, the single largest attribute-level skew in the panel; this shift is partly convergence rather than uniform amplification, since GPT-4o Mini favours female-authored content under \textit{popular} and \textit{engaging} ($-6.6$ and $-7.0$ points) but switches to favouring male authors under \textit{informative} and \textit{controversial} ($+1.7$ and $+7.3$ points), converging in direction with Claude and Gemini, who favour male-authored content under every objective ($+2.9$ to $+14.2$ points). The objectives that lean hardest on post content are also the ones under which political moderates lose the most visibility.

\subsection{Demographic Skew by Context Level}

\begin{figure}[t]
\centering
\includegraphics[width=.7\textwidth]{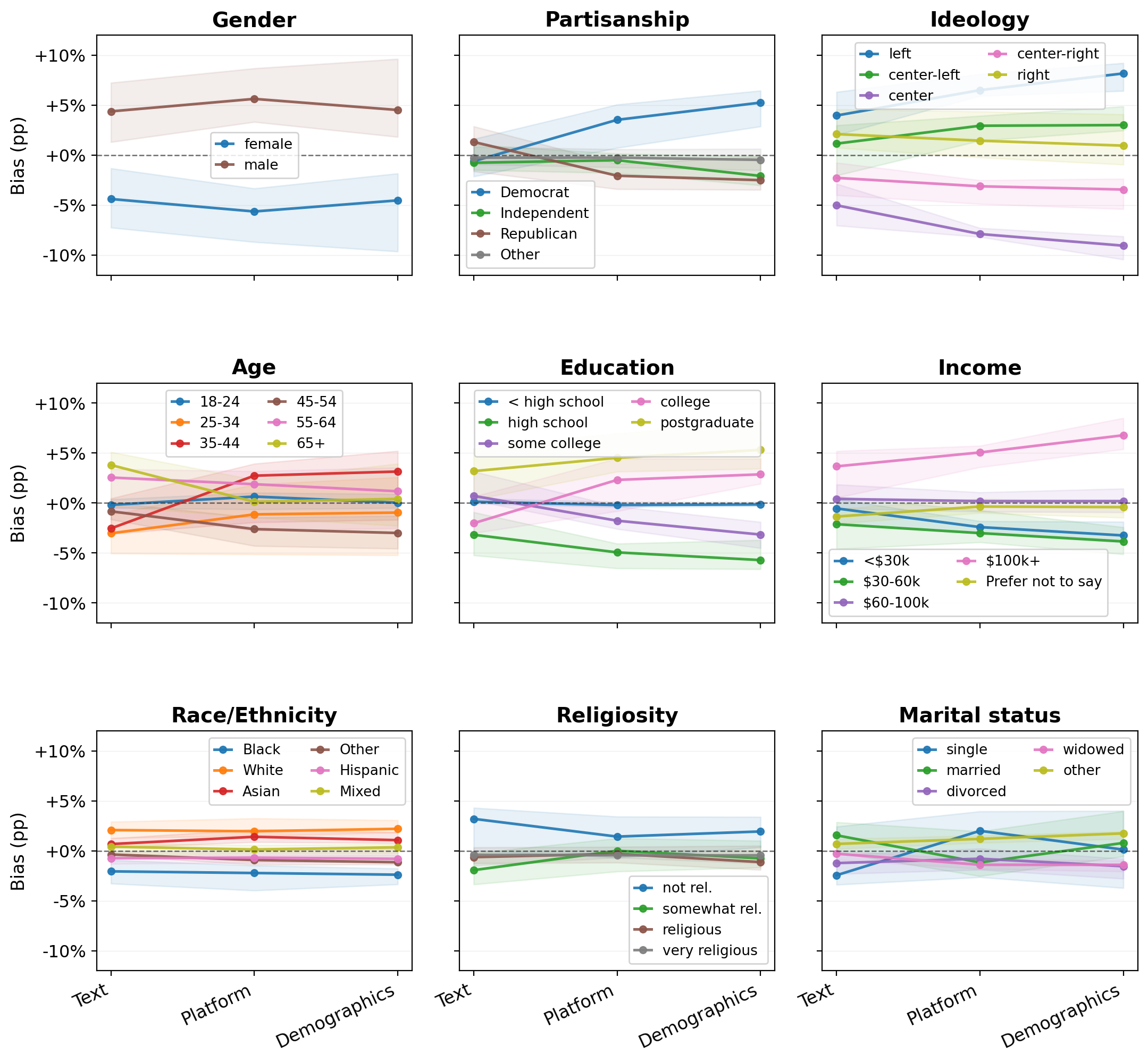}
\caption{Median demographic skew, with IQR across models and prompt styles, under \textit{text}, \textit{platform}, and \textit{demographics} conditions, disaggregated by category. Complements the aggregate text-vs-platform-vs-demographics comparison in Fig.~\ref{fig:demo-context}.}
\label{fig:demo-context-bygroup}
\end{figure}

Fig.~\ref{fig:demo-context-bygroup} shows the text-to-platform-to-demographics pattern disaggregated by category. Ideology exhibits the strongest and most consistent amplification: left-leaning over-representation grows from $+4.0$ percentage points under \textit{text} to $+8.2$ under \textit{demographics}, and centrist under-representation deepens from $-5.0$ to $-9.1$ points, while the right-leaning category is a partial exception, declining from $+2.1$ to $+1.0$ points as context is added. Partisanship shows a similar but starker pattern: Republican-authored content moves from over-representation ($+1.3$ points under \textit{text}) to under-representation ($-2.5$ under \textit{demographics}), while Democrat over-representation grows from near zero to $+5.3$ points. Education and income move in step with each other: high-school under-representation deepens from $-3.2$ to $-5.7$ points and the highest income bracket's over-representation grows from $+3.7$ to $+6.8$. Race, by contrast, is comparatively stable across all three context levels (e.g., Black under-representation moves only from $-2.1$ to $-2.4$ points), consistent with the near-diagonal clustering of racial categories in Fig.~\ref{fig:demo-context}.

\subsection{Where Demographic Skew Enters the Pipeline, by Prompt Style}

\begin{figure}[t]
\centering
\includegraphics[width=.8\textwidth]{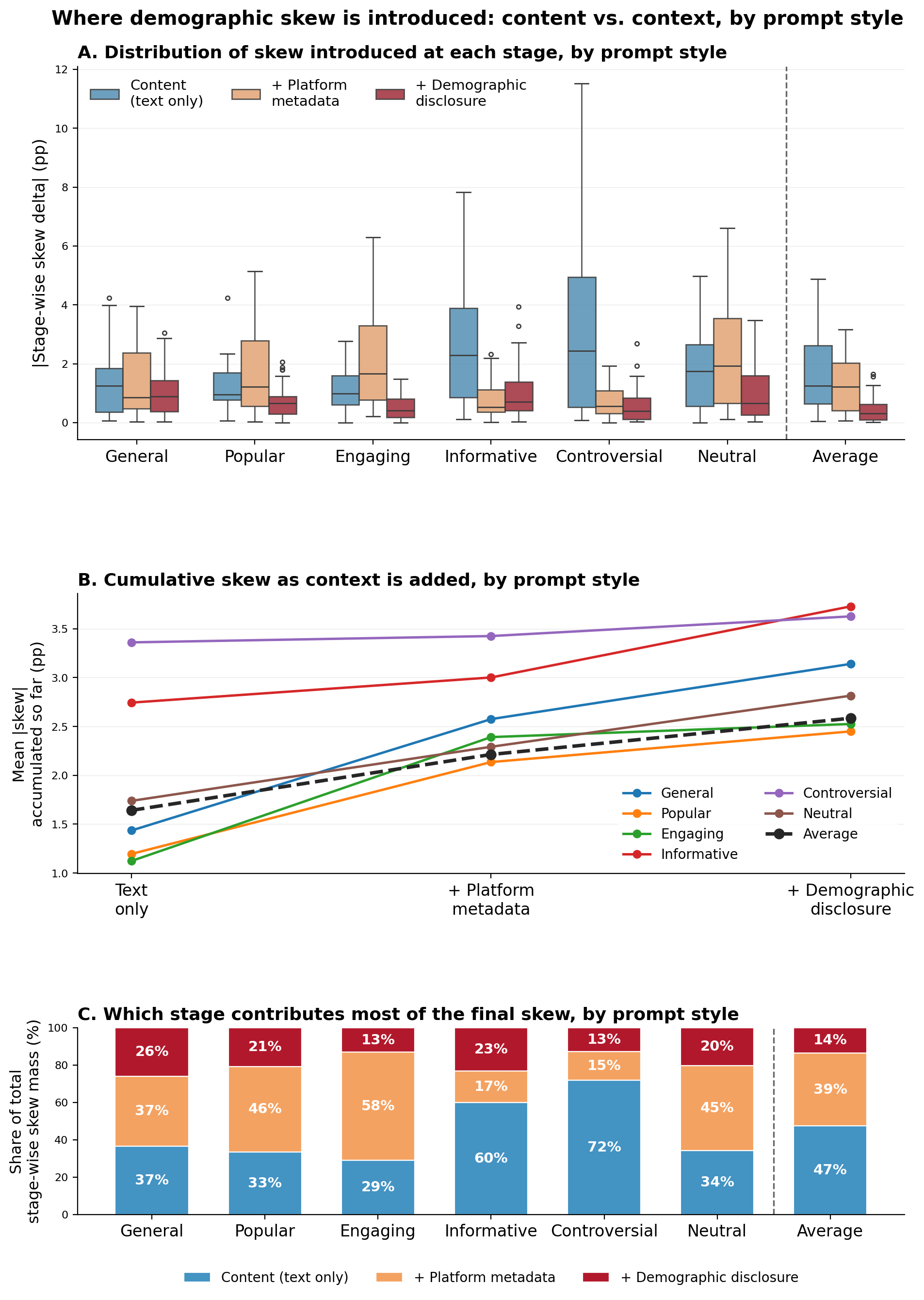}
\caption{Directional demographic skew decomposed into three additive pipeline stages, by prompt style (\textit{Average} pools all six styles). For each category in Fig.~\ref{fig:demo-context}, skew under \textit{demographics} is decomposed as skew under \textit{text} plus the increment from adding \textit{platform} metadata plus the increment from adding \textit{demographics} on top of \textit{platform}; the three terms sum exactly to the final skew. (\textit{A}) Distribution of $|$stage-wise skew delta$|$ (percentage points) across categories. (\textit{B}) Mean $|$skew$|$ accumulated so far at each stage (the level, not the delta), i.e., how much bias is present in total by that point in the pipeline. (\textit{C}) Each stage's share of a style's total stage-wise skew mass.}
\label{fig:bias-stage-decomposition}
\end{figure}

Fig.~\ref{fig:bias-stage-decomposition} decomposes the text-to-platform-to-demographics pattern of Fig.~\ref{fig:demo-context} by prompt style, and shows that its ``demographic disclosure adds little'' finding is an average over two different regimes. Under \textit{controversial} and \textit{informative}, content selection alone already accounts for $72\%$ and $60\%$ of each style's total stage-wise skew (panel \textit{C}), leaving comparatively little for platform metadata and demographic disclosure to add: adding platform metadata displaces categories by only $0.7$--$0.8$ points on average under these two objectives, against a text-only base of $2.7$--$3.4$ points. Under \textit{engaging}, \textit{popular}, and \textit{neutral}, by contrast, content alone accounts for only $29$--$34\%$ of total skew, and platform metadata does most of the remaining work ($45$--$58\%$), displacing categories by $1.6$--$2.3$ points on top of a much smaller ($1.1$--$1.7$-point) text-only base. Across the six styles, this text-only base and the displacement from adding platform metadata are strongly negatively correlated ($r=-0.81$): objectives that already skew heavily from content alone leave less room for platform signals to add further skew, and vice versa.

Despite these different starting points, panel \textit{B} shows that total skew converges by the final stage. \textit{Informative} and \textit{controversial}, the two most content-skewed objectives at \textit{text}-only, remain the two most skewed overall once demographics are disclosed ($3.7$ and $3.6$ points), but \textit{general}, \textit{engaging}, \textit{neutral}, and \textit{popular} close most of the gap over the two subsequent stages, ending between $2.4$ and $3.1$ points. \textit{Informative} in fact overtakes \textit{controversial} by the final stage, despite trailing it through \textit{platform}, because it continues to accumulate skew at each successive stage rather than plateauing after \textit{text}.

\subsection{Single-Block Conditional-Logit Results and Full-Joint Robustness Check}

Conditional-logit models that account for the fixed 100-post choice sets (Section~\ref{sec:methods}) show that the content preferences documented in the main text persist after the measured content characteristics are considered jointly. \textit{News \& Social Concern} remains the strongest positive topic predictor across all three providers ($b=0.88$--$1.03$), while text length is strongly positive ($b=0.40$--$0.60$). Sentiment polarity is negative ($b=-0.28$ to $-0.10$), indicating a preference for more negatively valenced posts, and subjectivity is positive ($b=0.12$--$0.28$). Posts containing mentions are consistently penalised. The selection profile is therefore not reducible to a single feature or to simple correlations among topic, length, and affect: the models systematically distinguish among several dimensions of expression even when ranking the same candidate pools.

\begin{figure}[t]
\centering
\includegraphics[width=\textwidth]{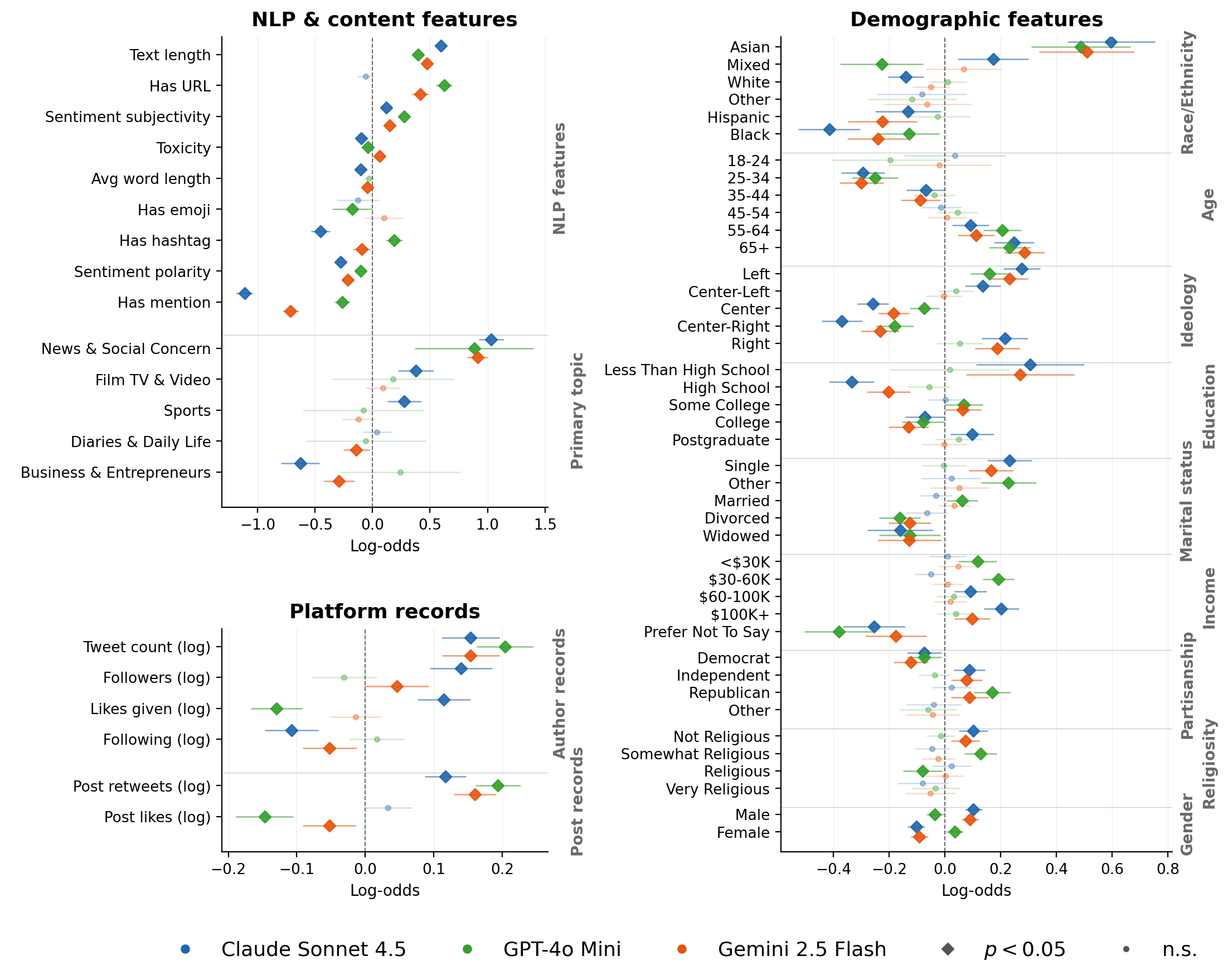}
\caption{Conditional-logit coefficients ($b$) in the \textit{text}-only condition, estimated by exact conditional MLE stratified by the trial pool shown to the model (Section~\ref{sec:methods}). Left panels show NLP/content features (top) and platform records (bottom); the right panel shows demographic features. Platform and demographic variables in this figure are observed characteristics of the posts and authors but were not supplied to the LLM in the \textit{text}-only condition. Features are $z$-scored; bars show 95\% CIs. Filled diamonds indicate $p<0.05$.}
\label{fig:reg-panels}
\end{figure}

Section~\ref{sec:demographic}, in the main text, reports headline attenuation statistics from the demographic-plus-content specification (demographic and content features fitted jointly, both fit on the \textit{text}-only condition so that neither model uses information the LLM did not observe; Section~\ref{sec:methods}; main-text Fig.~\ref{fig:rq2-demo}\textit{B}). This section summarizes those statistics, then isolates platform's own incremental contribution as a separate robustness check, using the full-joint specification (demographic, content, and platform features fitted jointly) even though platform metadata is never shown to the model in the \textit{text}-only condition. Among the 78 provider-category combinations significant in the single-block model, the median attenuation once content alone is added is 32.8\%, with 3 sign flips and 25 (32\%) losing significance. Adding platform characteristics on top of an already content-adjusted model shrinks the remaining associations by only a further 1.1\% at the median (among the 53 still significant after content controls alone; 0.8\% across all 78), with zero further sign flips and zero further losses of significance: platform metadata adds essentially no further confounding beyond content, consistent with the content-based mechanism (topic, length, sentiment, and style) documented throughout this section. The race asymmetry noted in the main text is already present in the single-block model and is simply unchanged once content, and then platform, are added. Full per-category coefficients are in the supplementary \texttt{demo\_attenuation\_after\_content\_controls.csv} (single-block vs.\ demographic-plus-content) and \texttt{demo\_attenuation\_platform\_increment.csv} (demographic-plus-content vs.\ full-joint) files.

\subsection{Demographic Correlations}

\begin{figure}[h!]
\centering
\includegraphics[width=0.85\textwidth]{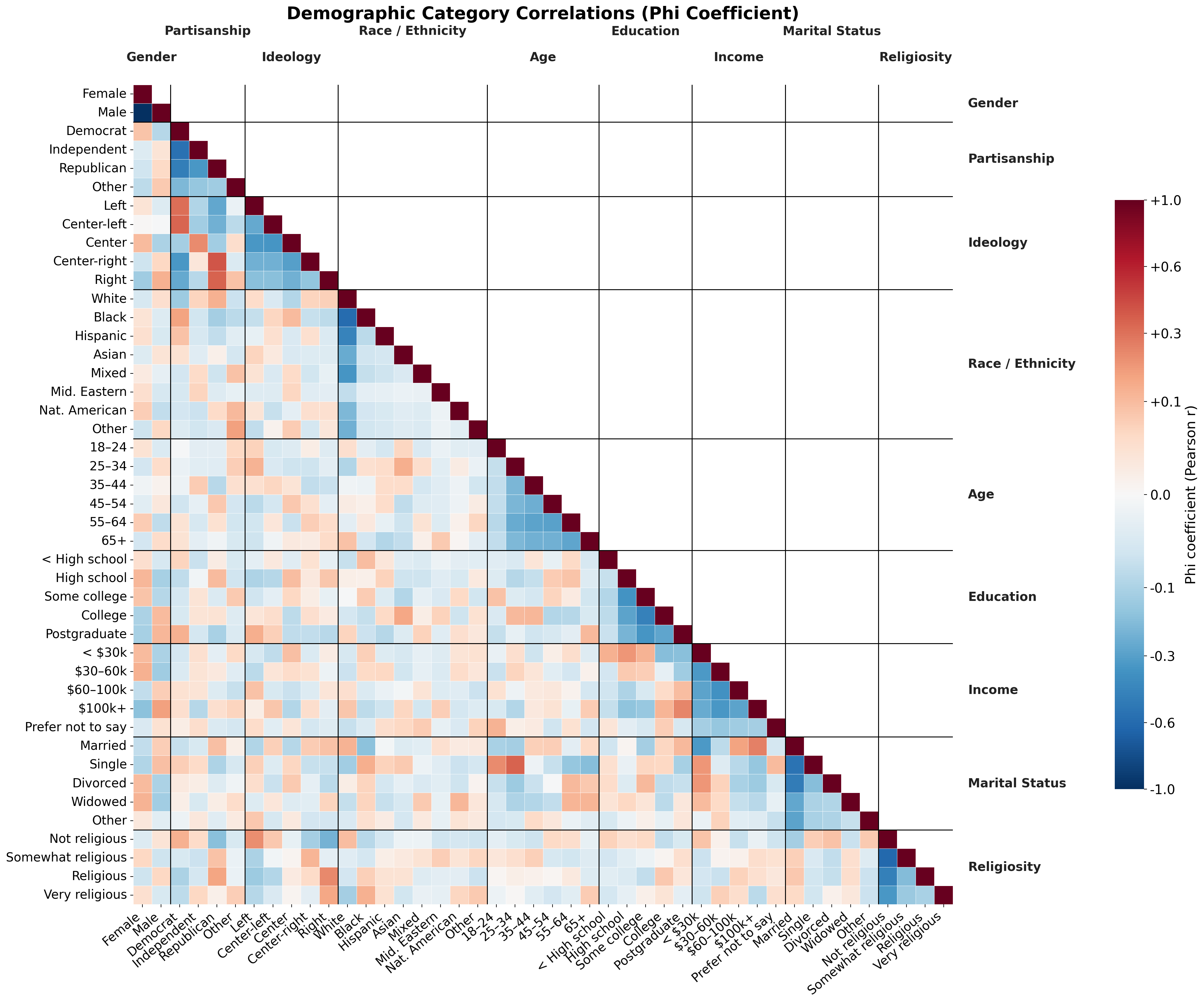}
\caption{Phi-coefficient (Pearson $r$ on binary indicators) between all 44 demographic category dummies ($N = 1{,}000$ YouGov panel authors). Blue = positive co-occurrence; red = negative/mutual exclusion. The strongest off-diagonal clusters are the partisanship--ideology alignment (Democrat with Left/Center-left; Republican with Right/Center-right), older age co-varying with married and religious, and education co-varying with higher income.}
\label{fig:demo-phi}
\end{figure}

Figure~\ref{fig:demo-phi} shows the phi-coefficient (Pearson $r$ on binary indicators) between all 44 demographic category dummies ($N = 1{,}000$ authors). The strongest off-diagonal positive correlations are between partisanship and ideology: Democrat co-occurs strongly with Left and Center-left, and Republican with Right and Center-right, confirming that the two attributes capture largely the same underlying political dimension in this panel. Older age groups (55--64, 65$+$) correlate positively with being married and with higher religiosity. Education and income are positively associated: College and Postgraduate categories co-occur with the \$60--100k and \$100k$+$ income brackets. Negative (red) off-diagonal entries largely reflect within-attribute mutual exclusion (e.g., Democrat vs.\ Republican, or different age groups).

\subsection{Demographic Inference Accuracy}

\begin{figure}[t]
\centering
\includegraphics[width=\textwidth]{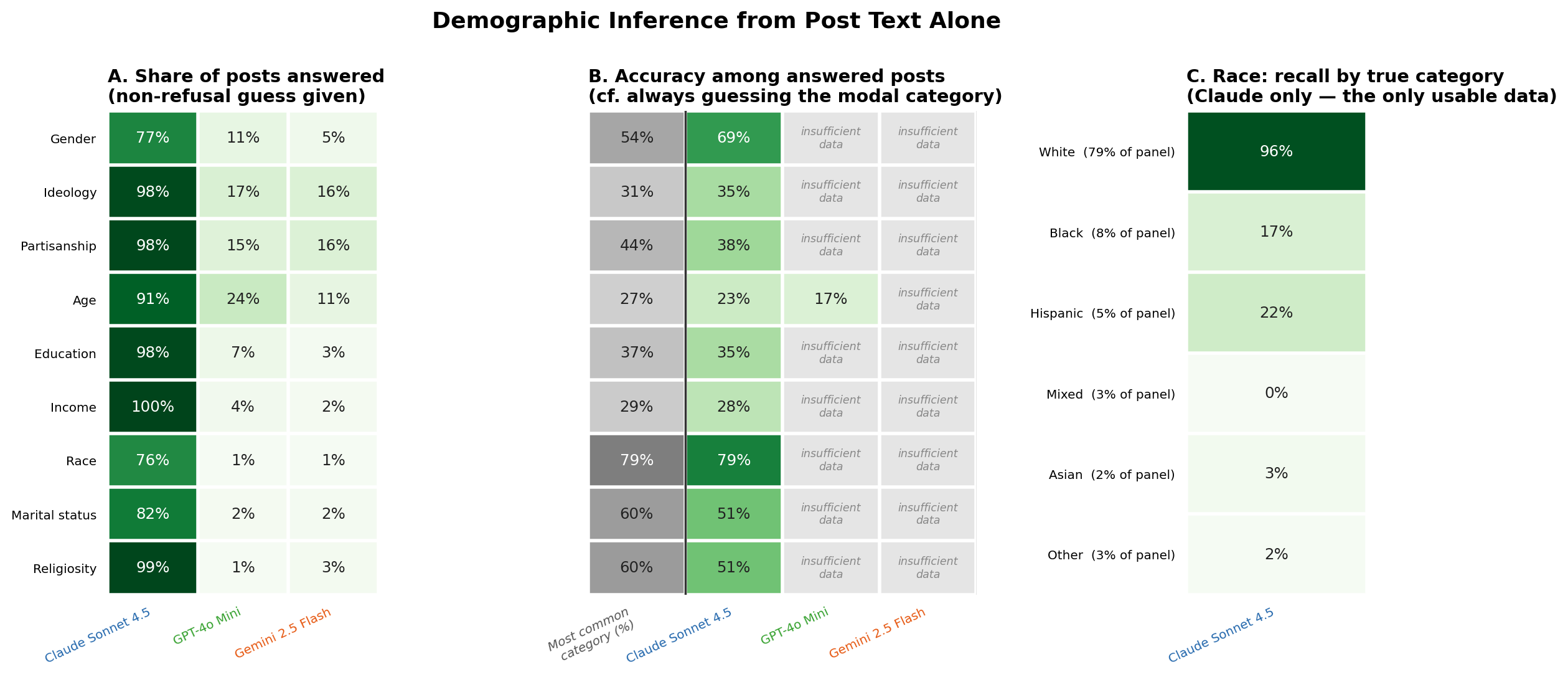}
\caption{Demographic inference from post text alone. (\textit{A}) Share of posts each provider gave a specific (non-refusal) guess for, by attribute. (\textit{B}) Accuracy of that guess among answered posts, against a reference column showing the share of authors already in that attribute's single most common category (the accuracy achievable by always naming that one category, ignoring the post). Cells built from fewer than 20\% of posts are marked \textit{insufficient data} rather than plotted. (\textit{C}) For race, and Claude only (the only provider with usable data), recall broken down by each true category, showing that the aggregate 79\% accuracy in Panel \textit{B} reflects near-perfect recall on White authors (96\%) and near-chance recall on every other group.}
\label{fig:demo-inference}
\end{figure}

Referenced from \textit{Limitations}: a supplementary test asked each provider to infer an author's demographics from post text alone, independent of and run separately from the recommendation task (Fig.~\ref{fig:demo-inference}). The reference column against which accuracy is judged is a harder, more meaningful bar than guessing uniformly at random among categories of very unequal size: the accuracy achievable by ignoring the post entirely and always naming an attribute's single most common category (e.g., 79\% White for race).

Panel \textit{A} shows why most of Panel \textit{B} has to be read with caution: Claude answers the large majority of posts for every attribute (75--99\%), but GPT-4o Mini and Gemini refuse for 76--99\% of posts depending on attribute. Cells in Panel \textit{B} built from fewer than 20\% of posts are marked \textit{insufficient data} rather than plotted, since an accuracy computed on that small, self-selected remainder (the posts a provider happened to feel confident enough to answer) is not a representative estimate.

Even where Panel \textit{B} \emph{is} estimated on enough data, the accuracy gains over that reference column are modest to nonexistent. Claude's race accuracy (79\%) is nearly identical to the 79\% already achievable by always naming the panel's most common category, White. Panel \textit{C} confirms this directly by breaking that 79\% down into recall for each true race category: Claude correctly identifies White authors 96\% of the time, but Black authors only 17\%, Hispanic 22\%, and Mixed, Asian, and Other authors 0--3\%. The aggregate accuracy is high only because the panel is dominated by White authors; the model shows essentially no ability to distinguish the other five categories from one another. Gender (69\% vs.\ 54\%) and ideology (35\% vs.\ 31\%) show a modest genuine edge over the reference column; accuracy is at or below it for the remaining six attributes. Overall, this test finds little evidence that any provider performs genuine fine-grained demographic inference from post text beyond exploiting category base rates.

\end{document}